\documentclass[useAMS,usenatbib,letterpaper]{mn2e}
\usepackage[totalwidth=480pt,totalheight=680pt]{geometry}
\usepackage{graphicx,amssymb,amsbsy}
\usepackage{times}
\usepackage{ gensymb }
\usepackage{amsmath}
\usepackage{verbatim}
\usepackage[T1]{fontenc} 
\usepackage{aecompl}
\usepackage{color}

\input psfig

\newcommand{\be}{\begin{equation}}
\newcommand{\ee}{\end{equation}}
\def\lta{\,\raise 0.3 ex\hbox{$ < $}\kern -0.75 em
 \lower 0.7 ex\hbox{$\sim$}\,}
\def\gta{\,\raise 0.3 ex\hbox{$ > $}\kern -0.75 em
 \lower 0.7 ex\hbox{$\sim$}\,}

\newcommand{\apj}{ApJ}
\newcommand{\apjs}{ApJS}
\newcommand{\pasp}{PASP}
\newcommand{\mnras}{MNRAS}
\newcommand{\procspie}{Proc. SPIE}
\newcommand{\apjl}{ApJ}

\title[Inclination Oscillations for Exoplanet Orbits] 
{Oscillations of Relative Inclination Angles\\ 
in Compact Extrasolar Planetary Systems} 
\author[Becker \& Adams]{Juliette C. Becker$^1$ and Fred C. Adams$^{1,2}$\\
$\,$\\ 
$^1$Astronomy Department, University of Michigan, Ann Arbor, MI 48109\\
$^2$Physics Department, University of Michigan, Ann Arbor, MI 48109 } 

\begin{document} 

\date{February 2015} 

\pagerange{\pageref{firstpage}--\pageref{lastpage}} \pubyear{2015}

\maketitle

\label{firstpage}

\begin{abstract}

The Kepler Mission has detected dozens of compact planetary systems
with more than four transiting planets. This sample provides a
collection of close-packed planetary systems with relatively little
spread in the inclination angles of the inferred orbits. A large fraction of the observational sample contains limited multiplicity, begging the question whether there is a true diversity of multi-transiting systems, or if some systems merely possess high mutual inclinations, allowing them to appear as single-transiting systems in a transit-based survey. 
This paper
begins an exploration of the effectiveness of dynamical mechanisms in
exciting orbital inclination within exoplanetary systems of this class.
\textcolor{black}{
For these tightly packed systems, we determine that the orbital inclination angles are not spread out appreciably
through self-excitation.
} 
In contrast, the two Kepler multi-planet systems with additional non-transiting planets are susceptible to oscillations of their inclination angles, which means their currently observed configurations could be due to planet-planet interactions alone. We also provide constraints and predictions for the expected transit duration variations (TDVs) for each planet. \textcolor{black}{ In these multi-planet compact Kepler systems, oscillations 
of their inclination angles are remarkably hard to excite; as 
a result, they tend to remain continually mutually transiting 
(CMT-stable). We study this issue further} by augmenting the planet masses and determining the enhancement factor required for oscillations to move the systems out of transit. The oscillations of inclination found here inform the recently suggested dichotomy in the sample of solar systems observed by Kepler.

\end{abstract}


\section{Introduction} 
\label{sec:intro} 

The Kepler mission has discovered a large number of compact extrasolar
systems containing multiple planets that can be observed in transit
\citep{lissauer,batalha}. Roughly forty of these such systems have
four or more planets. The inventory of these four-plus planet systems
includes mostly super-Earth sized planets, which have radii $R_P =
2-5 R_{\earth}$ and orbital periods in the range 1 -- 100 d.
Moreover, the orbital periods of the planets within a given system are
regularly spaced (roughly logarithmically uniform in period or
semimajor axis). Because all of the planets were observable by Kepler
at their times of discovery, these systems have an additional
stringent dynamical constraint: they must have retained a relatively
narrow spread in their orbital inclination angles. On the other hand,
orbital inclination can often be excited in close-packed planetary
systems. The goal of this paper is thus to explore the oscillations of
orbital inclination within solar systems of this class. Excitation of
inclination can be driven by a variety of mechanims, incluing unseen
additional companions, perturbations from stellar encounters in
clusters \citep{adams2001,gdawg}, and self-excitation through
interactions among the observed planets. This paper focuses on this
latter mechanism.

Slight deviations from true coplanarity in these systems (e.g., as
observationally supported in \citealt{rowe, kepler11}; and others)
allow for the possibility of oscillations in the inclination angles of
the planetary orbits, e.g., due to secular interactions between the
planets \citep[see also][]{55cancri_secular}.  If such oscillations were common, and had sufficient
amplitude, then not all members of a solar system could be seen in
transit at every epoch. As a result, multi-planet systems would
display evidence for ``missing'' planets, i.e., exceptions to the
(roughly) logarithmically even spacing of orbits that are often observed.
The ubiquity of this class of exoplanetary systems places strong
constraints on both their architectures and dynamical histories (see
also \citealt{chiang}). We note that the inclination angle oscillations for Jupiter and Saturn in our own solar system are large enough to periodically move the orbits out of a mutually transiting configuration. 

Statistical analyses of the Kepler system architectures suggest that
there could exist two distinct populations of planetary systems
\citep{ballard14,tim14}, namely, a population with single-transiting
planets and an additional population of multi-planet systems.  The
existence of these two distinct populations could
be explained by either two true distributions of solar systems (e.g.,
created by two different formation histories) or a single distribution
in which some systems exhibit a high degree of scatter in orbital
inclination angles.  Excitation of inclination in nearly coplanar
systems could shift some planets out of a transiting configuration,
thereby leading to the population of single-transit systems. In this
case, the single-transit systems would be a subset of the
multi-transiting group rather than a distinct population.

This paper explores possible oscillations of the inclination angles in
compact extrasolar systems. The measured planetary radii
$R_P = 2 - 5 R_{\earth}$ imply planetary masses $M_P = 4 - 30 M_{\earth}$, 
where we use \textcolor{black}{
a conversion law based primarily on the
probabilistic mass-radius relationship derived in \cite{2015arXiv150407557W}:
\begin{equation}
\frac{M}{M_{\earth}} \sim \rm{Normal} \left( \mu = 2.7 \left( \frac{R}{R_{\earth}}\right)^{1.3} , \sigma = 1.9 \right)
\end{equation}
where $M$ refers to the mass of a body, $R$ its radius, and this expression represents a $r^{1.3}$ scaling law with a normal distribution of scatter due to potential planetary composition variation. 
The Wolfgang relationship describes the a distribution of the potential masses for planets in the range $R_P = 1.5 - 4 R_{\earth}$. Since a small number of planets in our sample lie outside these bounds, we supplement the Wolfgang relation in two ways: for planets with radii $R_P < 1.5 R_{\earth}$, we
supplement with the rocky relation from \cite{lauren}; for planets with radii $R_P > 4 R_{\earth}$, we determine starting density using the Wolfgang relation, then add a scatter and choose a radius anomaly to account for varying core masses and inflation due to thermal effects \citep{radanom}. Of the 208 planets in our sample, only 9 fall above the regime described by the Wolfgang relation. 
}
 With
relatively large masses and close proximity, planet-planet
interactions can be significant. On the other hand, these planetary
systems orbit relatively old stars (with ages of $\sim1-6$ Gyr,
weighted toward the lower end of this range; see \citealt{basri}), so
that they are expected to be \textcolor{black}{dynamically} stable over $\sim$Gyr time scales. These systems are also generally non-resonant. These
considerations --- significant interactions coupled with long-term
stability and non-resonance --- suggest that the planetary systems are subject to
secular interactions
\citep{md}. In the present context, we are interested in secular
oscillations of the inclination angles of the orbits. If such
oscillations have sufficient amplitudes, the resulting spread of
inclinations angles in the system will sometimes be large enough that
not all of the planets can be seen in transit.  When observed in such
a configuration, the system will appear to have gaps in the regular
spacing of planetary orbits that these systems usually exhibit.  The
goal of this paper is to understand the amplitude of self-excitation
of inclination angle oscillations and provide limits on transit
duration variations, an observable with amplitude directly related to
inclination evolution over time, for observed Kepler systems with no
unseen companions. This analysis will allow future observations of transit
durations for these systems to inform the presence of massive outer
companions in these systems.

We note that spreads in the inclination angles can be produced by a
variety of astronomical processes. This work will focus on secular
oscillations of the inclination angles by the compact solar system
planets themselves (with semi-major axes $a\lta0.5$ AU). Future work
will focus on the effect of possible additional bodies in the outer
part of the solar system (where $a\approx5-30$ AU), roughly analogous to the giant planets in our outer Solar System. 

We stress that oscillations of inclination angles are not rare.
Within our Solar System, for example, the orbital inclinations of
Jupiter and Saturn oscillate with a period of about 51,000 years and
an amplitude of about $1^\circ$ (see Figure 7.1 in \citealt{md}). The
inclination angles of the two orbits coincide every half period
(25,500 years), so that an observer oriented in that plane would see
both planets in transit at that epoch. However, the amplitude of the
oscillation is sufficient to move both planets out of transit for an
appreciable fraction of the secular cycle.

\textcolor{black}{This paper focuses on the case of self-excitation of inclination
angles for Kepler systems with four or more planets, where the secular
dynamics of such systems are considered in Section 3. An analysis of
the observed compact, mutually transiting systems is presented in
Section \ref{mcmethod}, which shows that the systems are consistently mutually
transiting over time.  An orbital architecture that is continually
mutually transiting is denoted here as CMT-stable (which should not be
confused with dynamical stability). We consider a generalized class of
systems in Section 3.2, and study compact systems which have been
discovered to host an additional non-transiting planet in Section 3.3
(where these systems are shown to be more active).}
We also compare these results with numerical simulations in Section
\ref{sec:numerical}. 
Section \ref{sec:tranduration} presents observables for the compact Kepler systems discovered to
date; specifically, the transit durations are predicted to vary and the magnitude of these variations
are determined. In Section \ref{sec:mcs}, we study the stability of the observed Kepler systems by considering how the predicted oscillation amplitudes would vary if planet masses are scaled upward: the systems are found to be remarkably \textcolor{black}{dynamically} stable. The paper concludes, in Section \ref{sec:conclude},
with a summary of our results and a discussion of their implications,
as well as a statement on our plans for future work.

\section{Secular Theory for Inclination Angles}
\label{sec:secular} 

To evaluate the behavior of mutual inclination for these isolated
systems, we apply Laplace--Langrange secular theory \citep{md}. This 
formalism allows the use of the long-period terms of the disturbing
function to describe orbital motion over many secular periods. 

\subsection{Review of the Theory} 

We expand to second order in inclination and eccentricity, and first
order in mass. With this expansion, inclination and eccentricity are
decoupled, so we can write the disturbing function as a function of inclination alone:
\be
\mathcal{R}^{\rm{(sec)}}_{j} = n_{j}a_{j}^{2} 
\Biggl[ \frac{1}{2} B_{jj} I_{j}^{2} 
+\sum_{k=1}^{N} \left( B_{jk} I_{j}I_{k} \cos{(\Omega_{j} -\Omega_{k})}\right)   
\Biggr]\,,
\ee
\noindent where $j$ is the planet number, $n$ is the mean anomaly, $I$ is the 
inclination, $\omega$ is argument of pericenter,
and $\Omega$ is the longitude of the ascending node. The coefficients $B_{ij}$ are defined by 
$$
B_{jj} = -n_{j} \Biggl[ \frac{3}{2} J_{2} 
\left( \frac{R_{c}}{a_{j}} \right)^{2} - \frac{27}{8} J_{2}^{2} 
\left( \frac{R_{c}}{a_{j}} \right)^{4} - \frac{15}{4} J_{4}^{2} 
\left( \frac{R_{c}}{a_{j}} \right)^{4} 
$$
\be
\qquad \qquad  + \frac{1}{4} \sum \frac{m_{k}}{M_{c} + m_{j}}
\alpha_{jk} \bar{\alpha}_{jk} b^{(1)}_{3/2}(\alpha_{jk}) \Biggr] \,,
\label{bmatrixdiag} 
\ee 
and 
\be
B_{jk} = n_{j} \left[ \frac{1}{4}\frac{m_{k}}{M_{c} + m_{j}}
  \alpha_{jk} \bar{\alpha}_{jk} b^{(1)}_{3/2}(\alpha_{jk}) \right]\,,
\label{bmatrixoff} 
\ee
where $J_{2}$ and $J_{4}$ describe the oblateness of the central star (which we set to be $=0$ in all our analysis), $R_{c}$ is the stellar radius, $m_{k}$ indicates the mass of the $k$th planet, $M_{c}$ denotes the mass of the central star, $\alpha_{jk}$ denotes the semi-major axis ratio $a_{j}/a_{k}$, and $\bar{\alpha}_{jk}$ denotes the semi-major axis ratio $a_{j}/a_{k} < 1$. The quantities $b_{3/2}^{(1)}$ is 
the Laplace coefficient, which is defined by  
\be
b_{3/2}^{(1)} = \frac{1}{\pi} \int_{0}^{2\pi} 
\frac{\cos{\psi}\ d \psi}{(1-2 \alpha \cos{\psi} + \alpha^{2})^{3/2}}\,,
\ee
(as given in \citealt{md}). All of the coefficients $B_{jk}$ can be considered as frequencies that describe the
interaction between each pair of planets, and are elements of the
matrix denoted as \textbf{B}.  This application of
secular theory allows us to evaluate the problem analytically, but 
neglects higher-order terms. In this formulation, the only terms in
the disturbing function are those that do not depend on the mean
longitudes, as we assume that the short-period terms average out over long timescales.  The coefficient matrix \textbf{B} describes
inclination evolution. Solving for the matrix
elements of \textbf{B} allow us to determine the time evolution of
inclination.

The matrix \textbf{B} defines an eigenvalue problem \citep{md}, where
the eigenvalues describe the interaction frequencies between any pair
of planets. The eigenfrequencies of this matrix, denoted here as
$f_{i}$, along with the eigenvectors ${\cal I}_{jk}$, can be used to
describe the time evolution of the system. Given the matrix
\textbf{B}, we can solve for the eigenvalues and eigenvectors using
standard methods.  With these quantities specified, we also need the
initial conditions to specify the full solution for the time evolution
of the inclination angles $I_j$ and the angles $\Omega_j$.  It is
convenient to transform the dependent variables according to 
\be
p_j = I_j \sin \Omega_j \qquad {\rm and} \qquad 
q_j = I_j \sin \Omega_j \,,
\ee
so that the solutions take the form 
\be
p_j (t) = \sum_{k=1}^N I_{jk} \sin (f_k t + \gamma_k) 
\label{ptime} 
\ee
and 
\be
q_j (t) = \sum_{k=1}^N I_{jk} \cos (f_k t + \gamma_k) \,,
\label{qtime} 
\ee
where the phases $\gamma_k$, along with the overall amplitudes, are
determined by the initial conditions. The quantities $I_{jk}$ are
eigenvectors, where we use the standard (but awkward) notation such
that the first index $j$ specifies the planet number and hence the 
components of the eigenvector and the second index $k$ runs over the
different eigenvectors. It is also useful to define normalized
eigenvectors ${\cal I}_{jk}$ and corresponding scaling factors $T_k$
such that
\be
I_{jk} = T_k {\cal I}_{jk} \,.
\ee
The initial conditions then specify the scaling factors through 
the expressions 
\be 
p_j (t=0) = \sum_{k=1}^N T_k {\cal I}_{jk} \sin \gamma_k 
\ee
and
\be 
q_j (t=0) = \sum_{k=1}^N T_k {\cal I}_{jk} \cos \gamma_k \,.
\ee 

The scaled eigenvectors $I_{jk}$ (which conform to the system's
boundary conditions), the eigenvalues $f_{k}$, and the phases
$\gamma_{k}$ are sufficient to specify the time evolution of the
orbital inclination of each body in the system, i.e.,
\be 
I_{j}(t) = \sqrt{\left[ p_j(t) \right]^2 + 
\left[ q_j(t) \right]^2 } \,,
\label{inc_excite} 
\ee 
where the solutions $p_j(t)$ and $q_j(t)$ are given by equations
(\ref{ptime}) and (\ref{qtime}).  Implicit in this solution is the
linear dependance on the interaction coefficients (the matrix elements
given by equations [\ref{bmatrixoff}]). From this
solution, we note that the inclination evolution has a linear dependance on mass
ratio and a second order dependence on the semi-major axis ratio
between the planet in question and each planet exterior to its orbit.

\section{Inclination Oscillations due to Self-Excitation} 
\label{sec:selfexcite} 

The compact Kepler systems with four or more planets are tightly
packed systems with minimal mutual inclinations. From this population,
it appears that planets in multi-transiting systems generally have
non-null values of mutual impact parameter, and subsequently
inclination \citep{rowe}. Systems with non-null mutual inclinations
exhibit non-parallel angular momentum vectors, allowing the
possibility of excitation in inclination and other orbital
elements. To test the magnitude of these excitations, we take the
population of all Kepler systems with four or more transiting planets as examples of compact, multi-body, transiting systems. 
\textcolor{black}{We obtain our data from the NASA Exoplanet Archive\footnote{http://exoplanetarchive.ipac.caltech.edu}, updating system parameters when newer values have been found \citep[such as in the case of Kepler-296;][]{2015arXiv150501845B}.}

\begin{figure*} 
\centering
\includegraphics[width=6.5in]{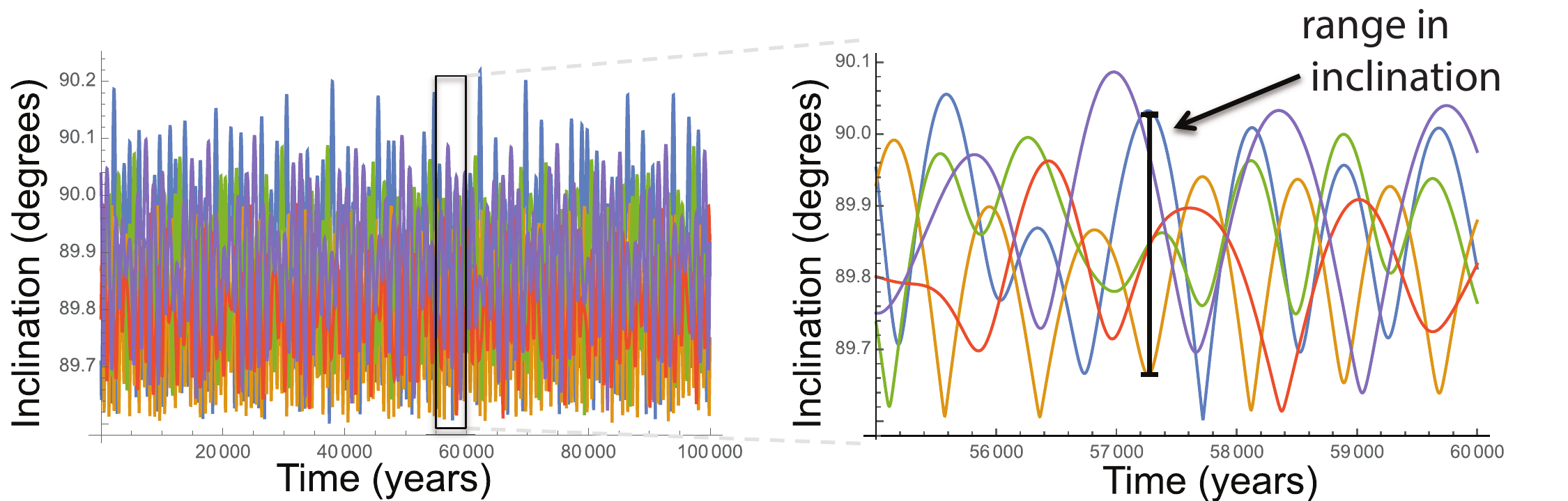} 
\caption{Plotted here are the inclination evolutions of five roughly coplanar 
planets, with initial conditions drawn from the priors of Kepler-256. Although the inclinations of the planets
generally stay within a plane, there is also instantaneous variation,
which manifests as \textcolor{black}{a range in the inclinations of} plane of planets. This variation may lead
planets to be knocked out of a transiting configuration. The mutual inclination, shown on the right panel, changes as planets precess, meaning that the width of the plane containing all the planets oscillates over time. } 
\label{fig:map} 
\end{figure*}

There are observational biases inherent in the Kepler systems, \textcolor{black}{as a photometric transit survey is by definition more likely to find systems with low mutual inclinations and aligned argument of pericenters \citep{RH2010}. The Kepler multi-planet systems are likely more aligned and more compact in inclination plane width than an 'average' system, but the sample found by Kepler is} 
representative of the type of system we would expect to see
from photometric transit surveys such as Kepler \citep{keplermission1}, K2 \citep{k2mission}, and TESS \citep{tessmission}. 
\textcolor{black}{
It is not currently clear, however, whether the Kepler multi-planet systems that we do see in transit are CMT-stable or if we are catching them at a lucky moment in which all planets appear to be in transit. This differentiation is important because the former possibility describes a much less dynamically active system than the latter. 
}
To test the
stability against exciting planets out of the transiting plane, we
used the secular theory described in Section \ref{sec:secular} to
numerically evolve each system in the Kepler multi-planet sample for
several secular periods. This procedure results in a measure of the spread in impact parameter $\Delta
b(t)$ (see below). \textcolor{black}{We also compute the probability that the system is mutually transiting, marginalized over all trials and realizations in our simulations.} If $\Delta b(t) <2$ for an entire secular period \textcolor{black}{and the probability of all planets transiting simultaneously for a random time-step in a random realization of the system is high ($P(\rm{transit}) > 0.85$)} then the system is said to be \textcolor{black}{CMT-}stable in a
transiting configuration. Note that the condition of being \textcolor{black}{CMT}-stable against oscillating planets out of transit is much more confining that being \textcolor{black}{dynamically} stable against planet ejection. For a given Kepler system, we can use a
Monte Carlo analysis to evaluate $\Delta b(t)$ not just once, but many
times, with starting orbital elements for each realization selected
from observationally motivated priors.  For parameters that have been
measured (for transiting systems, the radius of the planet $r_{p}$ and
the semi-major axis $a_{p}$, and sometimes the inclination $I_{p}$ or
eccentricity $e_{p}$), we draw each planet's orbital element from a
normal distribution with mean and standard deviation determined from
observations. For orbital elements not measured, we draw a value from
priors summarized in Table \ref{priors}.

\textcolor{black}{Observationally measured inclinations have been fit from photometric light curves, and for these planets there is a degeneracy between angles over 90\degree and under 90\degree. The literature reports inclination angles as $< 90\degree$, so when we use a literature measurement, we choose a value not only from that planet's measured posterior but also choose its orbit to fall above or below the midplane of the star with equal probability. 
For planets without measured inclinations, we choose a plane width from a Rayleigh distribution with width 1.5\degree \citep{fab2010}, subject to the constraint that all planets must be transiting. This choice of distribution follows work done by \cite{2009ApJ...696.1230F, lissauer, fangmar, ballard14}. In these recent works, Rayleigh distributions with varying widths are used to describe the size of the plane containing the planets. The value we use here, 1.5\degree, is within the range suggested by the work of \cite{fab2010}.
}

\textcolor{black}{We note that the
argument of the ascending node is not necessarily independent of the
value of inclination angle as assumed here. As planetary systems
evolve to attain nonzero inclination angles, modeled here by a
Rayleigh distribution, the nodes will evolve into some other
distribution, which should be characterized in future work.}

\begin{table}
\centering
\label{table1}
\centerline{\bf Orbital Element Distributions}  
\centerline{$\,$} 
\begin{tabular}{cc}
\hline 
\hline
Parameter & Prior \\ 
\hline
$\omega$& uniform on  $(0^\circ, 360^\circ) $\\
$\Omega$& uniform on  $(0^\circ, 360^\circ) $\\
$e$	& uniform on  $(0, 0.1) $\\
$I$	& \textcolor{black}{Rayleigh distribution with width $\sigma = 1.5\degree$} \\
\hline
\hline
\end{tabular}
\vspace{0.25cm}
\caption{When orbital elements have not been measured observationally,
we draw their values randomly from the prior distributions summarized 
in this table. }
\label{priors} 
\end{table}    

Once we have the initial conditions for each Kepler system, we can evolve orbits as according to the secular theory described in Section \ref{sec:secular}. This must be done individually for each realization of initial conditions for each system. 

\subsection{Evaluating the Secular Behavior of the Compact, Multi-Planet Kepler Systems}
\label{mcmethod}

A tightly packed, roughly coplanar system of planets will trade
angular momentum as the system evolves (while keeping the total angular momentum vector of the system constant).  The magnitude of this
exchange determines the magnitude of the variations in orbital
elements of each body.  Equation (\ref{inc_excite}) describes the
inclination evolution for each body in a system.  Once the inclination
solutions for each planet in a system have been found using equation
(\ref{inc_excite}), a comparison between them (see Figure \ref{fig:map}, which illustrates how the mutual inclination can change over time)
yields a measure of the mutual inclination between all planets in the
system. This mutual inclination describes the width of the plane
containing all the planets.

\begin{figure} 
\centering
\includegraphics[width=3.2in]{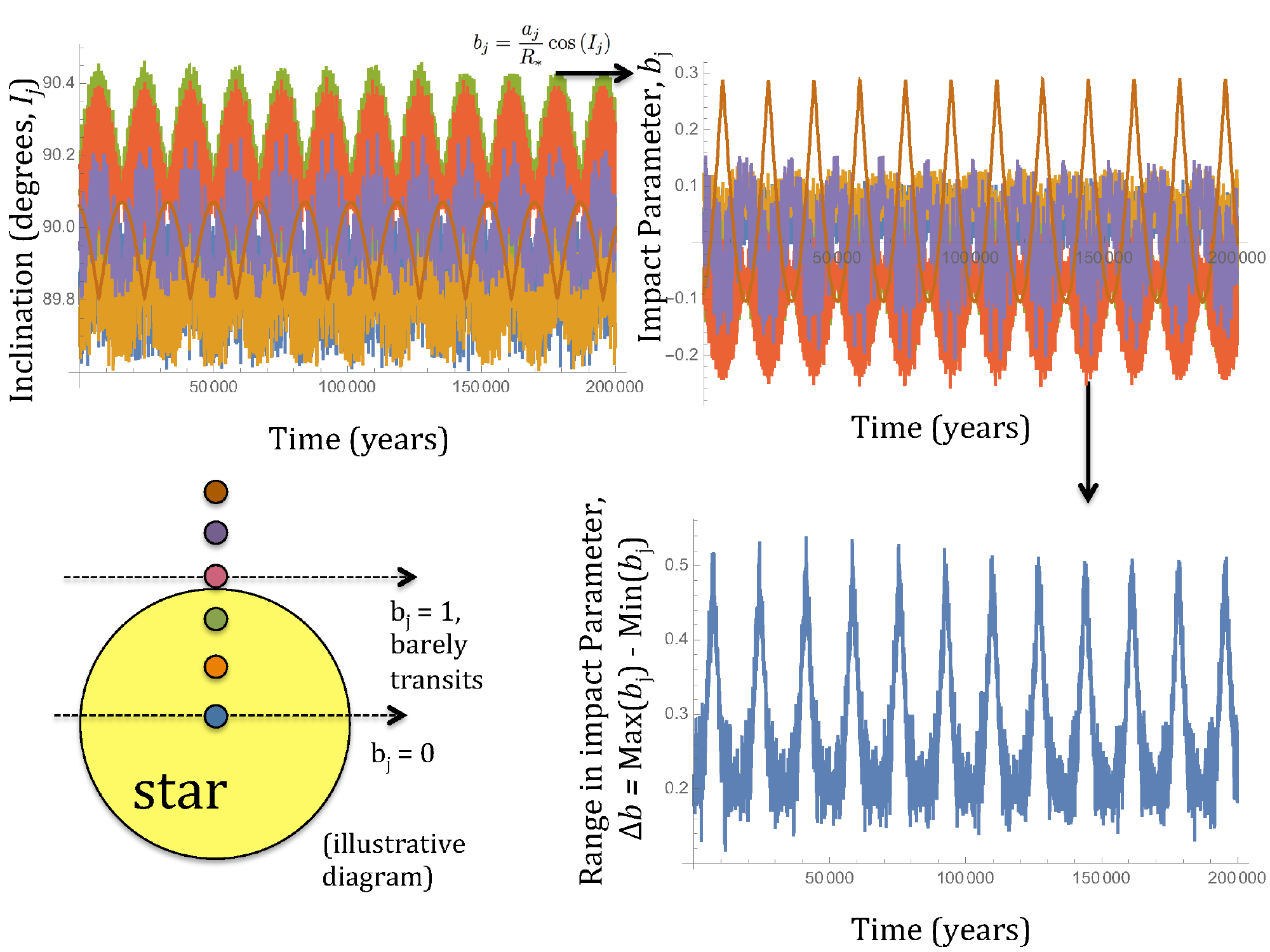} 
\caption{The parameterization of mutual impact parameter, as illustrated 
by test case Kepler-11. 
First, a plot of inclination for all bodies in a system
(upper left) is generated by solving equation (\ref{inc_excite}) with
the initial parameters of the system as boundary conditions. The semi-major axis dependency
is removed using equation (\ref{impact_parameter}), and the result,
impact parameter over time for each planet, is shown (upper
right). \textcolor{black}{The inclinations attained by each planet result in vastly different impact parameters due to the
differences in semi-major axis.} Planets closer to the star can attain
more inclination with less effect on their impact parameter. Finally,
the \textcolor{black}{range} between impact parameters is calculated (lower right) as
was done for mutual inclination in Figure 1. The result is a measure
of the \textcolor{black}{range} of the mutual impact parameter over time, $\Delta b(t)$. \textcolor{black}{As long as this width describes a plane that lies entirely within the limbs of the star, the planets will be CMT-stable.} }
\label{show_the_variables}
\end{figure}

As the condition for transiting is more
rigorous than approximate coplanarity (even as planets' inclinations
vary in concert, the planets with larger orbital separations are more
likely to cease transiting), we remove the dependence on orbital and stellar properties by working in terms of impact parameter, $b$, which is defined as: 
\begin{equation}
b_{j} = \frac{a_{j}}{R_{*}} \cos{(I_{j})}
\label{impact_parameter}
\end{equation}
where $j$ is planet number, $a$ is the semi major axis, $R_{*}$ the
radius of the central star, and $I$ the inclination. When $-1 < b_{j}
< 1$, planet $j$ will transit.  Using the analytic expression for inclination evolution (Equation \ref{inc_excite}), we can describe the long-term behavior of not only individual planets but the \textcolor{black}{range} of their
respective impact parameters.
The process of extracting the mutual impact parameter
$\Delta b$ is shown in Figure \ref{show_the_variables}.

Using this technique, we explored the evolution of orbits for the entire initial condition parameter space for each Kepler multi-planet system. For a given system, we conducted 4000 Monte Carlo trials for each Kepler system, resulting in
4000 realizations of $\Delta b(t)$, with different initial conditions drawn from the observational priors, supplemented with the values in Table \ref{priors}. This sample can be used to calculate
the mean \textcolor{black}{range of the} impact parameter over time for the Kepler system, as well as the width of the plane of planets in impact-parameter space.

\begin{figure} 
\centering
\includegraphics[width=3.3in]{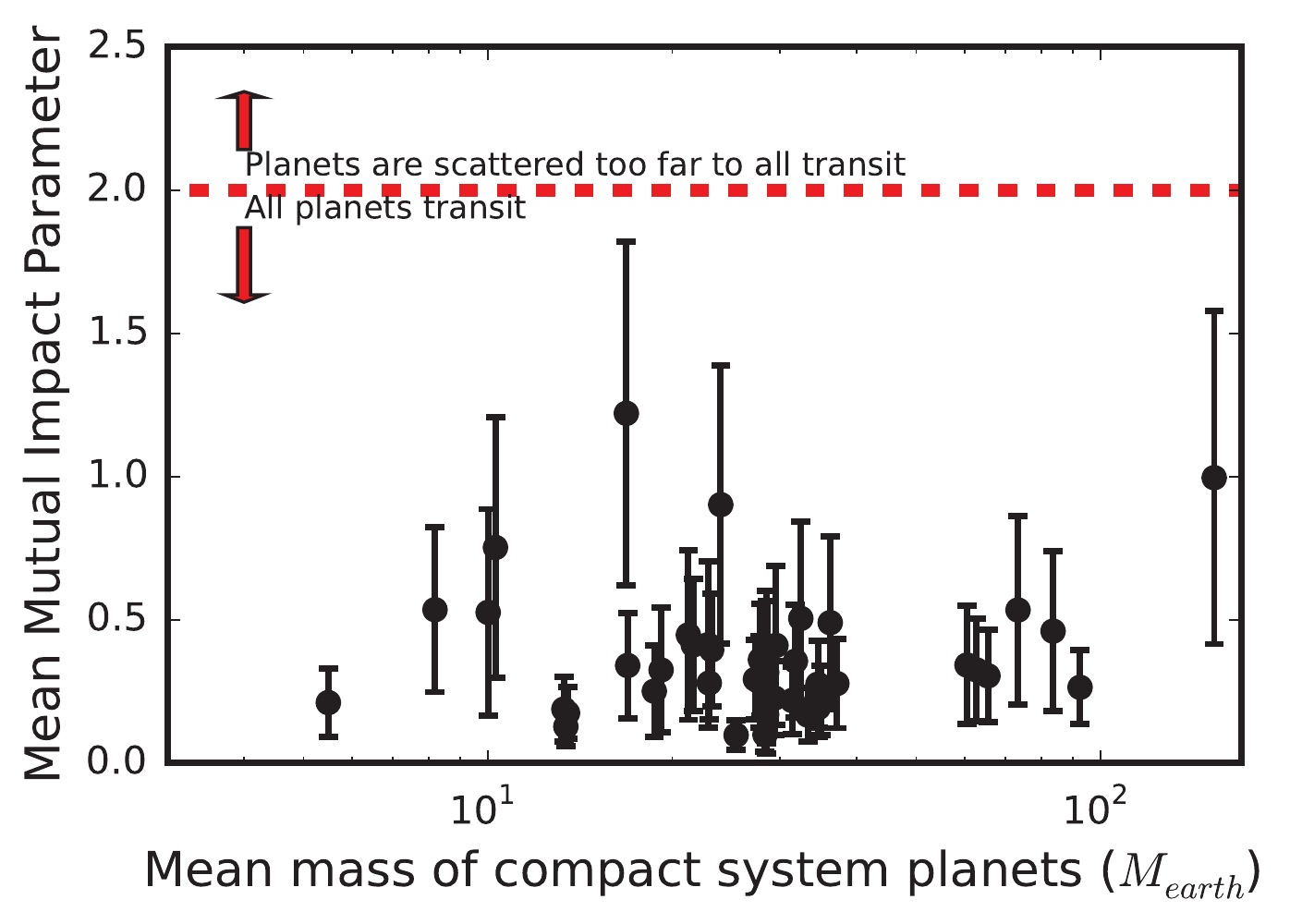} 
\caption{For each multi-planet \emph{Kepler} system, the parameters of  
the system were sampled 4000 times and evolved forward in time. The
resulting inclination angles for the planetary orbits were converted
to a mutual impact parameter (see text). The mean and scatter of these
values are plotted here for each system as a function of the total
mass of the transiting planets, given in earth masses. The dotted
horizontal line indicates the level above which it is not possible to
observe all the planets in transit. 
}
\label{one}
\end{figure}

We repeated this process of 4000 Monte Carlo trials for each for the 43 systems in our sample of multi-planet Kepler systems, resulting in a measure of the inclination evolution behavior for each system. 
Figure \ref{one} visualizes the results of these trials, where each point represents the mean mutual impact parameter for a different Kepler system. Mean mutual impact parameter is the typical width of the plane containing all planets in the system, and must be smaller than the diameter of the star for all planets to transit. An impact parameter plane width of \textcolor{black}{$\Delta b=2$}, marked on the plot, is the upper limit for all planets in the plane to be transiting.

For each point in Figure \ref{one}, the height of the point as compared to the transit limit ($\delta b=2$) corresponds to the width of the plane containing all the planets. The scatter (represented by error bars) corresponds to the width of the distribution due to the variations between realizations. 
For all systems, the projected plane containing the planets is much smaller than the diameter of the star, which means we \textcolor{black}{would expect to see all the planets in transit at for the majority of the secular history of the system}.

\begin{figure} 
\centering
\includegraphics[width=3.3in]{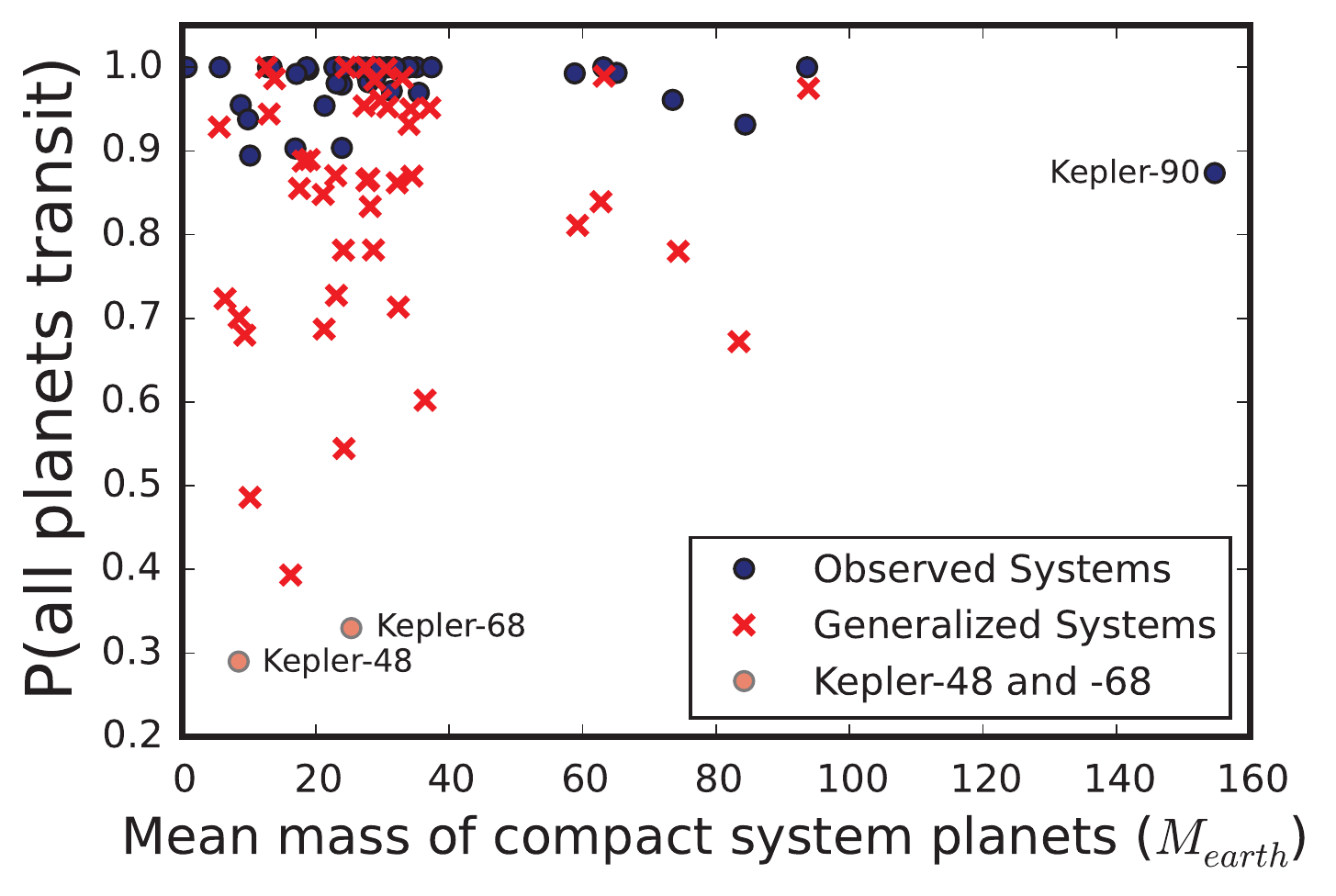} 
\caption{\textcolor{black}{For all realizations considered in Sections, \ref{mcmethod}, \ref{sec:DL}, and \ref{sec:nontransit}, we plot as circles the probability that a randomly chosen time-step from a randomly chosen realization will have all planets transiting along the line of sight to Earth. For the observed Kepler systems, all systems are mutually transiting more than 85\% of the time. This \textcolor{black}{result} indicates that statistically the observed Kepler systems are seen in transit an overwhelming majority of the time. The generalized systems, plotted as crosses, are mutually transiting a much lower fraction of the time, as are Kepler-48 and -68, the observed currently non transiting systems.}} 
\label{probcompare}
\end{figure}

\textcolor{black}{This parameterization represents the average behavior of each system over time. The plane width demonstrates how much range in impact parameter is normal for each architecture of system. However, we care about the transiting behavior of each system with respect to a single line of sight: that of the observer (Kepler) who originally identified the planets as mutually transiting. For example, it would be possible for a system's impact parameter range to be small enough for it to be possible for all planets to transit, but for the plane to be situated in such a location that only some planets transit. To understand how likely this is to happen, we plot in Figure \ref{probcompare} the mutual transit probability for each observed system as blue circular points. This probability is defined as the probability that a random time-step from a random realization, chosen from the sample of all 4000 realizations considered in the construction of Figure \ref{one}, will have all planets transiting along the line of sight to Earth. A probability of 1 would mean that the planets never left a transiting configuration in any time-step in any of our simulations, while a probability of 0 means that the system was never mutually transiting in any time-step in any realization. }

\textcolor{black}{Figure \ref{probcompare} shows that for the observed Kepler systems, all planets are expected to be transiting more than 85\% of the time. Indeed, for most systems the probability of mutual transit is even closer to 100\%. This demonstrates that not only do we expect the Kepler multi-planet systems to have plane widths small enough to potentially be transiting (Figure \ref{one}), the majority of the time they should maintain these transiting configurations with respect to our line of sight (Figure \ref{probcompare}).}

\textcolor{black}{From an analysis of the results in Figure \ref{one} and Figure \ref{probcompare}, it appears that while}
Kepler systems do excite mutual inclinations due to their dynamical
interactions with each other \textcolor{black}{(as their mutual impact parameters do change over time), the} magnitude of these
interactions are small enough that although an initially non-null
mutual inclination exists, it \textcolor{black}{remains, through the process of secular evolution,} smaller than the threshold necessary for
planets to not be observed in transit. \textcolor{black}{From this, we can state that the observed Kepler systems are generally CMT-stable.}

The Kepler systems with four or more planets do not exhibit
sensitivity to self-excitation of inclination due to dynamic
interactions between the inner, roughly coplanar planets. This result
indicates that self-excitation (in the mode considered here) is not a dominant
mechanism in knocking planets out a transiting plane and thereby creating
tightly-packed systems in which only some planets transit.

\textcolor{black}{It it important to note that the analysis of these observed Kepler system is limited by several factors: the measured mutual inclinations will be artificially low compared random systems drawn from the true distribution of planetary architectures, as these are systems with narrow enough ranges in inclination to be discovered in transit in the first place; the impact parameters of observed systems are likely artificially low due to the signal-to-noise bias against higher impact parameters; the deviation between measured planetary arguments of pericenter will also be artificially low \citep{RH2010}. These systems are not a representative sample of the true distribution of systems. As a result, the analysis presented here for the observed Kepler systems is not an analysis of the underlying planet population, but only of this particular class of heretofore discovered systems. }

\textcolor{black}{
\subsection{Inclination Oscillations in Generalized Kepler Systems}
\label{sec:DL}
The Kepler systems that we see are observationally biased in that they likely have unusually low mutual inclinations and aligned arguments of pericenter \citep{RH2010}. As we have shown in Section \ref{mcmethod}, the observed Kepler systems are remarkably CMT-stable in their transiting configurations. We are not simply lucky to see these systems in transit, merely viewing them at an opportune time: instead, we are seeing systems that will likely be consistently transiting over many secular timescales. 
The Kepler systems are indeed a special class of system. It would also be interesting to compare their behavior with that of generalized Kepler systems, with a wider range of starting orbital parameters.
}

\textcolor{black}{
To construct these systems, we repeat the following process for each Kepler system in our sample:}
\begin{itemize}
\item \textcolor{black}{Generate a compact planetary system based on the target Kepler system. To do this, we draw each orbital parameter from an inflated distribution, treating measured 3$\sigma$ errors as the width of our prior from which to draw orbital parameters. We convert radii to masses using the extended Wolfgang relation. }

\item \textcolor{black}{We evaluate the system for dynamical stability using the Hill-radii criteria outlined in \cite{fab2010}. We compute the separation between two orbits ($\Delta$) in terms of their Hill radii:
\begin{equation}
\Delta = (a_{out} - a_{in}) / R_{H}
\end{equation}}
\textcolor{black}{when the mutual Hill radius is given by:
\begin{equation}
R_{H} = \left( \frac{M_{in} + M_{out}}{3 M_{*}} \right)^{1/3} (a_{out} - a_{in}) / 2
\end{equation}
and for a system to be considered dynamically stable, $\Delta > 2 \sqrt{3}$ and for each pair of planets, $\Delta_{in} + \Delta_{out} > 18$ \citep{fab2010}.
\item If the system is dynamically stable according to these Hill arguments, we evolve the system and repeat the process for another set of starting parameters.
}
\end{itemize}
\textcolor{black}{Once this process is completed for each Kepler system, we have a sample of analog Kepler systems, which are based on the observed systems but no longer exactly the systems that we observe. This sample allows us to compute the mean mutual impact parameter over time, just as we did for the observed Kepler systems in the previous section. }

\begin{figure} 
\centering
\includegraphics[width=3.3in]{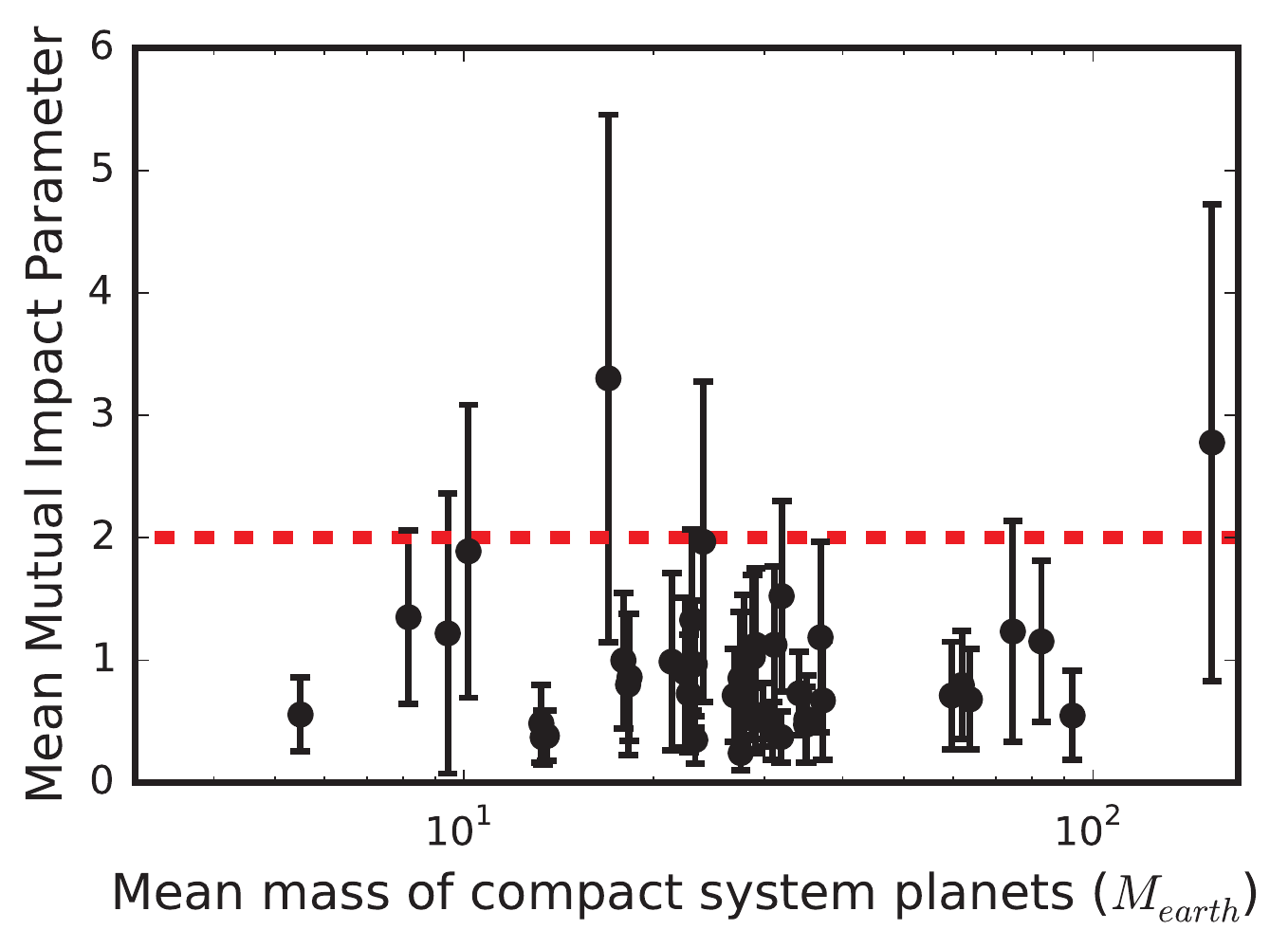} 
\caption{\textcolor{black}{For the generalized multi-planet \emph{Kepler} systems, the parameters of  
the system were sampled 4000 times and evolved forward in time, just as in Figure \ref{one}. The
resulting inclination angles for the planetary orbits were converted
to a mutual impact parameter (see text). The mean and scatter of these
values are plotted here for each system as a function of the total
mass of the transiting planets, given in earth masses. The dotted
horizontal line indicates the level above which it is not possible to
observe all the planets in transit. 
}} 
\label{one_v2}
\end{figure}

The result is shown in Figure \ref{one_v2}, which shows the same statistic plotted in Figure \ref{one} computed from the generalized Kepler systems. For these generalized systems, the \textcolor{black}{range of the} impact parameter over time is higher, suggesting that the Kepler systems we observe are a particularly CMT-stable subset of the dynamically possible compact systems that could exist. \textcolor{black}{Figure \ref{probcompare} shows as red crosses the mutual transit probability (over all time-steps and all realizations) for these generalized systems, demonstrating that the generalized systems spend significant amounts of time in non-mutually transiting configurations, as their plane widths imply they should. }

\subsection{Inclination Oscillations in Systems with Non-transiting Planets}
\label{sec:nontransit}

Long-term RV followup to systems with transiting planets has not only found masses for Kepler planets, but has also resulted in the characterization of additional, non-transiting companions to some transiting systems \citep{marcyrvsurvey}. Additionally, transit-timing variation analysis \citep{agol, holman} has both confirmed masses of planets and provided additional candidate planets \citep{cochran, hadden}.
The current state of these systems provides insight to their dynamical history: assuming that systems form from roughly coplanar protoplanetary disks, something in the evolution of these systems has resulted in sufficiently large spread in inclinations to prevent all planets from being seen in transit. 

As shown in Section \ref{mcmethod}, the
observed multi-transiting Kepler systems are \textcolor{black}{CMT-stable} against
self-perturbation (mutual inclinations excited by dynamical
interactions between the transiting planets). \textcolor{black}{Furthermore, the generalized Kepler systems are more likely than not to be seen in mutual transit}. For multi-planet systems
with some planets transiting and additional non-transiting companions,
something in the dynamical history of the systems has resulted in misalignment in inclination between the
planets. This effect could be explained in one of many ways: it could be due to a difference in formation
mechanism between the purely multi-transiting systems and the systems
with some planets outside the transiting plane; it could be due to some other perturbation, such as an as-yet undiscovered stellar or massive planetary companion (a possibility beyond the scope of this paper); or finally, it could be due to
the effect of self-excitation between all (known) planets in the system. 
Our analysis probes this final possibility, which would apply if all discovered planets (both those that are currently transiting and those that are currently non-transiting) in a system had started out roughly coplanar, in a potentially transiting configuration, and then through secular
interactions some planets had been perturbed out of the transiting
plane.

We can test this explanation for the currently observed misalignment of Kepler systems that have been
discovered to have multiple transiting planets and additional,
non-transiting companions using the same method that was used to evaluate the transit stability of the most tightly packed Kepler systems in Section \ref{sec:selfexcite}. Two examples of systems of this architecture are Kepler-48 and
Kepler-68. By starting the planets of these systems in
transiting configuration, we force the starting conditions to be a roughly coplanar disk containing all the planets. 

Kepler-48 \citep{kepler48, marcyrvsurvey} is a four planet
system with three inner transiting planets and one non-transiting
companion at more than 1 AU (a minimum mass 657 $M_{\earth}$ companion with a period of roughly a 980 day period).
Kepler-68 \citep{kepler68} is a three
planet system with two transiting planets and one non-transiting
planet, also outside of 1 AU (with a minimum mass of 0.95 $M_{jup}$ companion in roughly a 580 day period). 

To evaluate the transit stability of Kepler-48 and Kepler-68, we performed the same Monte Carlo
evolution described in Section \ref{mcmethod}, with all orbital
parameters drawn from observationally \textcolor{black}{constrained} priors except
inclination. 
\textcolor{black}{Though the true orbital inclination of the outer planets in the Kepler-48 and Kepler-68 systems is not known, we choose the orbital inclinations for the giant outer planets in each system by drawing a mutual inclination plane width from a Rayleigh distribution with a width of 1.5\degree \citep[from][which suggested a Rayleigh distribution width between 1.0\degree and 2.2\degree]{fab2010}. We constrain this choice of plane width such that the planets all start out mutually transiting, to mimic the starting conditions of the compact Kepler systems. With these starting conditions, we are probing what would happen to the observability of these systems over time, if they did start on feasibly observable architectures. 
}

\begin{figure} 
\centering
\includegraphics[width=3.3in]{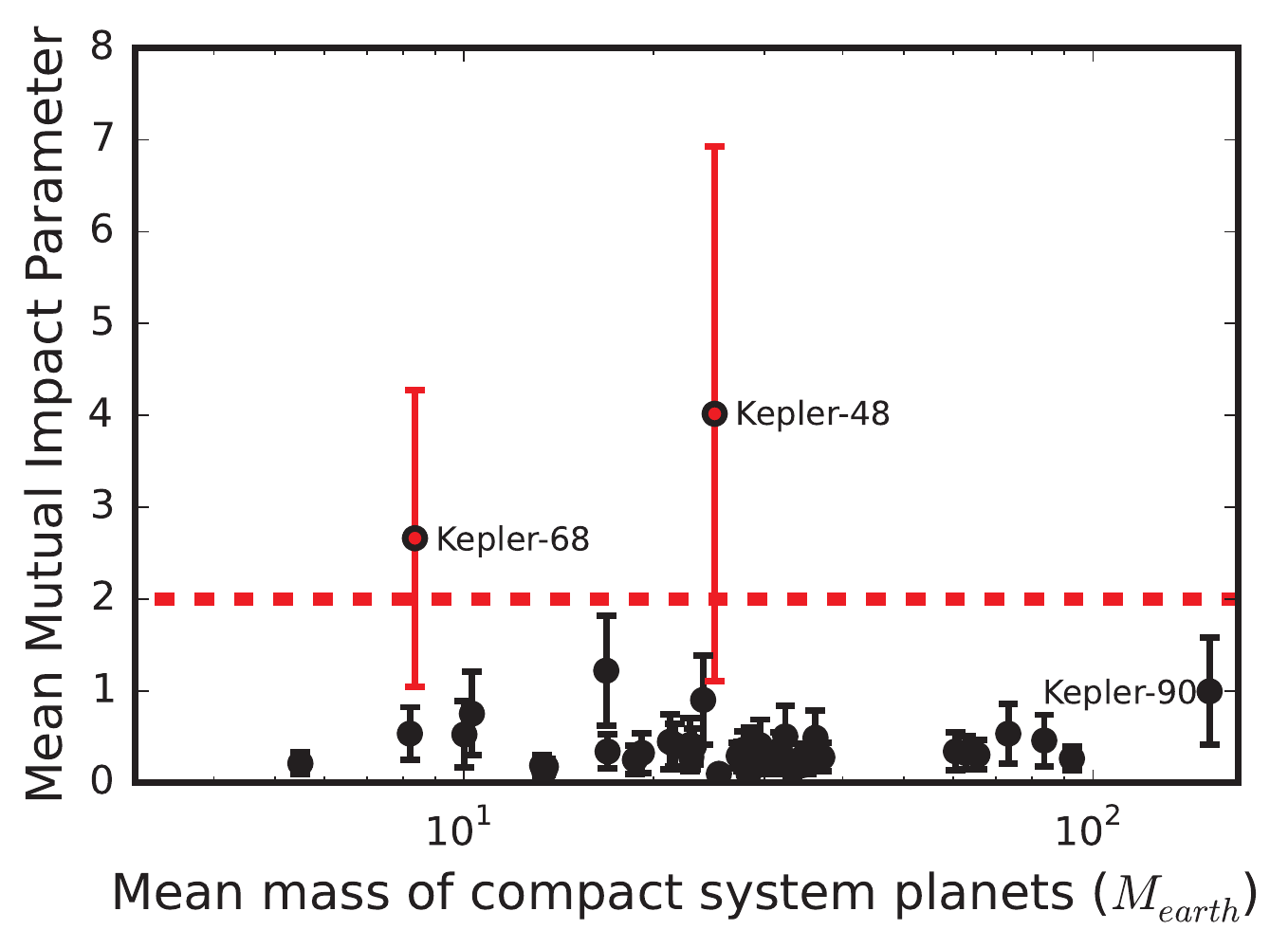} 
\caption{Kepler systems in which all discovered planets are transiting  
are plotted as black points (they correspond to the same data
presented in Figure \ref{one}), while Kepler systems where additional
non-transiting planets have been discovered are plotted as red points.
Kepler-48, marked \citep{kepler48, marcyrvsurvey} is a four planet
system with three inner transiting planets and one non-transiting
companion outside of 1 AU. Kepler-68 \citep{kepler68}, marked, is a
three planet system with two transiting planets within 0.1 AU and one
additional non-transiting planet at 1.4 AU.  }
\label{fig:nt}
\end{figure}

Through 4000 trials, Kepler-48 and Kepler-68 exhibited significantly more \textcolor{black}{range in their mutual} impact parameters than the other compact Kepler systems. Figure \ref{fig:nt} plots the behavior of Kepler-48 and Kepler-68 overlaid on the previous result for the compact Kepler systems. Kepler-68's mean mutual impact parameter is well above the limit for a mutually transiting system, while Kepler-48 spends about 60\% of its orbits in a transiting configuration (marginalized over starting parameters). 

We treat Kepler-48 and Kepler-68 as isolated systems. In other words, in our experiments, the only perturbation available to excite oscillations in inclination is that of the interactions between known bodies in each system. 
Thus, by generating the mean mutual impact parameter over one
secular period for these systems after they start in a transiting
configuration, we can make a statement about the amplitude of
self-excitation in these compact systems. As shown in Figure \ref{fig:nt}, both Kepler-48 and Kepler-68 would be expected to develop significant mutual inclinations that prevent all planets from being seen in transit purely through excite self-excitations
of inclination. \textcolor{black}{Figure \ref{probcompare} shows as salmon points the mutual transit probability for these two systems, confirming that it is unlikely that the magnitude of the secular interactions would allow these two planets to be seen in transit. }

\textcolor{black}{This result indicates that} even if these systems were to begin their secular evolution in a roughly coplanar configuration, they would be expected to self-excite sufficient oscillations to produce the current orbital state (where not all planets transit - we do not have sufficient limits o\textcolor{black}{n the} observed inclinations to make a stronger comparison). Kepler-48 and Kepler-68 are examples of systems that `make sense' dynamically: it is not required to add additional effects (such as a perturbing companion or stellar flyby) to their systems to explain their current non-transiting nature.
\textcolor{black}{It is important to note that the outer planets in these two systems are significantly external to the standard compact systems described in Section \ref{mcmethod}, which generally fell within 0.5 AU of their host star. Kepler-48 and -68 have outer companions at roughly 1.4 and 1.8 AU, respectively. It is possible that part of the reason for the activity of these systems is the lower transit probability of these outer companions, but the presence of Kepler-90 (which has an outer companion semi-major axis of roughly 1 AU) in the CMT-stable sample indicates that external companions do not ensure non-transiting configurations. }

\subsection{Comparison to Numerical Integrations} 
\label{sec:numerical}

\begin{figure} 
   \centering
   \includegraphics[width=3.4in]{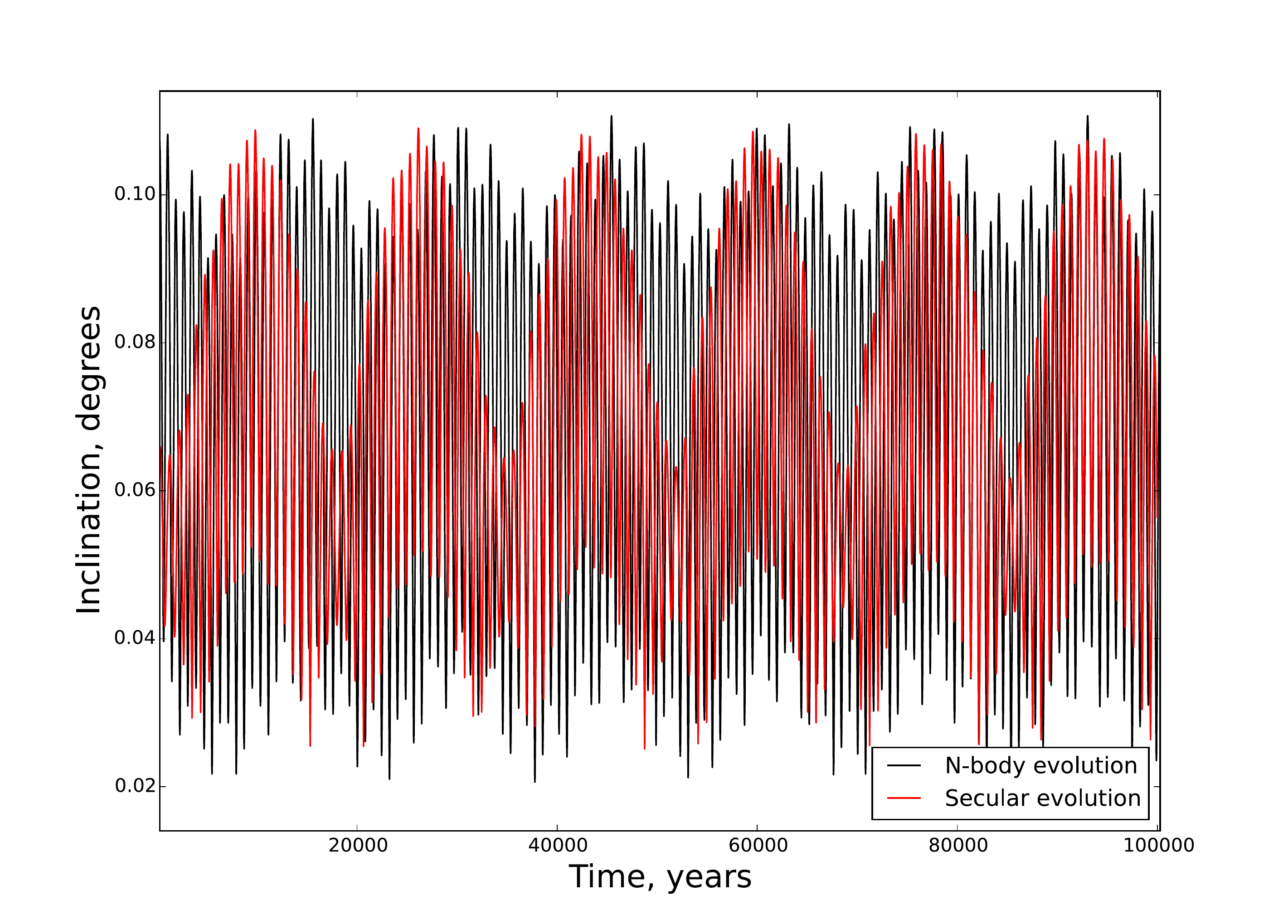} 
   \caption{An illustrative realization of Kelper-341b, with the result from the numerical N-body code \texttt{Mercury6} plotted in black and the secular theory evolution plotted in red.}
   \label{symp1}
\end{figure}

The discussion thus far has considered inclination oscillations as
described by second-order Laplace-Lagrange theory. Although the
amplitudes of the oscillations are small, so that the second order
theory is expected to be accurate, in this section we compare the
results to numerical simulations. These latter calculations, by
definition, include interactions to all orders. 

\begin{figure} 
   \centering

   \includegraphics[width=3.2in]{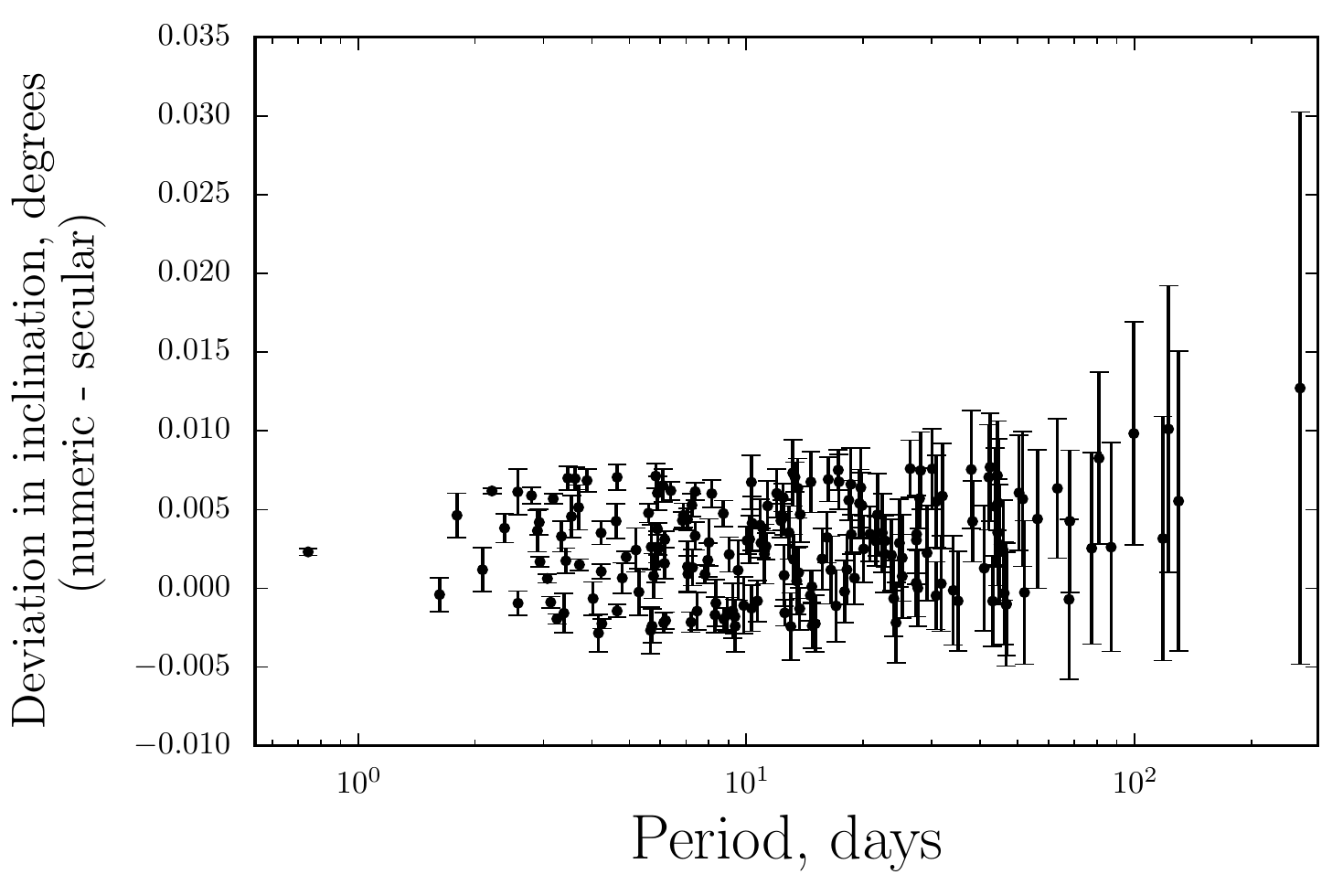} 
      \includegraphics[width=3.2in]{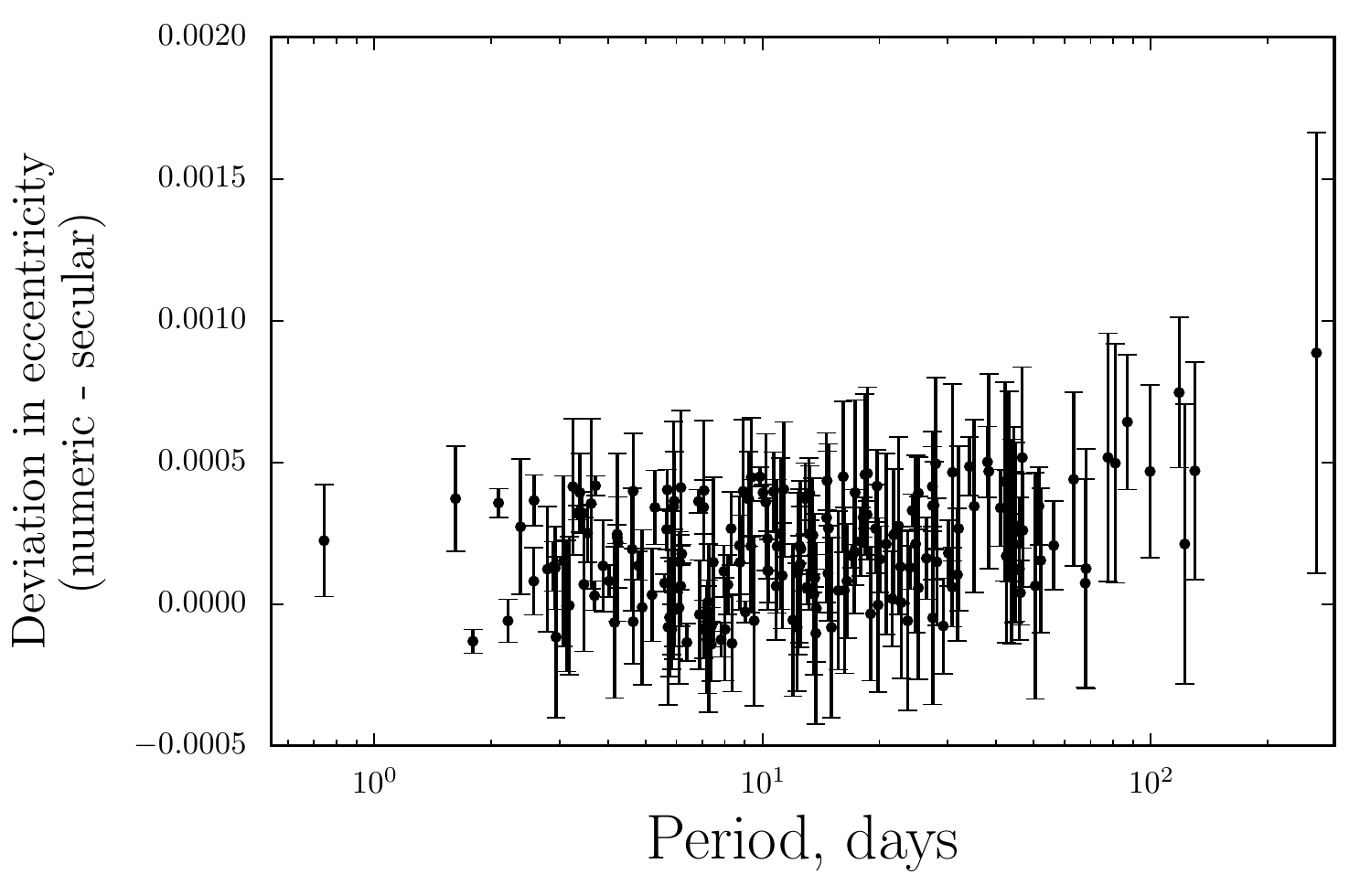} 
   \caption{A residuals plot of (\emph{top panel}) the deviation in inclination over several secular periods for each planet in our sample and (\emph{bottom panel}) the deviation in eccentricity for the sample sample of realizations. The averaged deviation in inclination between the numerical and secular methods is generally below 0.01$\degree$ for all planets.  }
   \label{symp} 
\end{figure}

For these compact systems, eccentricity and inclination are generally low, but to evaluate the error inherent in our second-order expansion, we evolved each compact system using hybrid symplectic and Bulirsch-Stoer integrator \texttt{Mercury6} (Chambers 1999). The numerical integrator should provide the effectively `right' answer, and significant deviations between the second-order theory and numerical results would indicate that second order secular theory is insufficient to describe the evolution of the orbital architectures. We compared 400 numerical N-body realizations with 400 secular evolutions (see the visualization of one realization of the comparison in Figure \ref{symp1}) to compute the deviations plotted in Figure \ref{symp}, which describe the mean deviation, in degrees, between secular theory and the numerical results. This comparison yielded a standard deviation of the difference in inclination angle obtained using secular theory and numerical results; this value was found to be less than 0.01\degree. 

For our use of second order second theory to be adequate for further analysis, we would want this variation between the numerical result and secular result to be much smaller than the threshold for significant inclination (which can cause a planet to become non-transiting). The planet in our sample with the largest semi-major axis and largest number of planets in the system, Kepler-11g, orbits a star with a radius $R_{*} = 0.0053$ AU. This planet would need to attain an inclination of 0.65$\degree$ out of the plane of the other planets to no longer transit. Planets with semi-major axes less than this value would need an even larger range of inclinations to be no longer seen as mutually transiting. Given that the typical deviation between the \texttt{Mercury6} numerical results and secular theory is less than 0.01$\degree$, the match between secular theory and N-body numerics is good enough to use the second order secular theory for these compact systems.  

We additionally note that although there is variation in the period of secular effects between numerical and second-order secular theory \citep{veras}, this does not affect our result, as we are concerned with the amplitude rather than period of inclination oscillations, and these amplitudes are well-predicted to a reasonable precision. If we were concerned with the exact period of secular effects, second-order Laplace-Lagrange theory would not \textcolor{black}{always} be sufficient. 

\textcolor{black}{
Finally, for completeness we note that the standard deviation of the
residuals between the secular and numerical results is not the only
measure of the difference (e.g., one could use the difference between
the {\it ranges} of inclination angles instead). In this case, however,
the differences between the two approaches is small: The differences 
would have to be nearly 100 times larger in order to change our main 
conclusion, i.e., that the Kepler compact systems remain CMT-stable.
}

\section{Transit Duration Variations}
\label{sec:tranduration} 

Oscillations of the orbital inclination angles, as described in
secular theory through equation (\ref{inc_excite}), result in planets
taking different paths across the face of the star as a function of
time.  These changing chords, in turn, result in the duration of the
planetary transit varying with inclination and hence with time. For
the case of vanishing eccentricity, we can write $\tau_{T}$, the time
from first to fourth contact (the transit duration) for a single
transit analytically (see \citealt{seager}), in the form 
\be
\tau_{T}(t) = \frac{P}{\pi} \arcsin \Theta \,
\ee
where we have defined the effective angle 
\be
\Theta \equiv \frac{R_{*}}{a} \left[  
\frac{(1 + r_{p}/R_{*})^2 - (a/R_{*}\cos^{2}i\textcolor{black}{)}}{1-\cos^{2}i} 
\right]^{1/2} \,,
\ee
where $P$ is the period of the planet, $a$ is its semi-major axis,
$R_{*}$ the radius of the central star, $r_{p}$ is the radius of the
planet, and $i$ is the inclination of the plane; note that the
inclination angle is a function of the time $t$ at which the duration
is evaluated (so that the duration will also be a function of time). We also assume that orbital elements are effectively constant during a single transit, but that variations occur from transit to transit. 
Substituting equation (\ref{inc_excite}) into this expression then yields a
measure of the transit duration, $\tau$, at any point during a
planet's secular evolution. The second order secular theory used in
this work computes motions with the evolution of inclination decoupled
from that of eccentricity, so the null eccentricity approximation for
extracting transit durations from our derived transit parameters is
sufficient.  A product of our stability study of the Kepler systems is
time series of $I(t)$ and subsequently $\Delta b(t)$. From these
expressions, we can compute the times series $\tau_{T}(t)$, evaluated
at each transit epoch for each planet in a system.

Thus far, observational study of secular TDVs has been limited by two main
factors: (1) the signature of TDVs caused by even massive planets is
generally small due to small yearly changes in inclination and
eccentricity, and (2) to find TDVs to good precision, the cadence of
photometric measurements must be high enough such that durations can
be extracted from individual transits.  Through TTVs can been used to
determine dynamical quantities of multi-planet systems with good success
\citep{agol, holman}, TDVs in multi-planet systems are generally as much as several order of
magnitude smaller in amplitude \citep[see, for example, Figure 4
  in][which demonstrates the difference in amplitude between a TTV and
  TDV signal for one system]{ttvking}. However, there has been recent
success measuring the amplitude of planetary TDVs
\citep{wasp3}. Since transit duration depends on the chord a planet
takes across its star in our line of sight and oscillating inclination
can directly change this chord, secular interactions exciting
inclinations will also lead to potentially observable transit duration
variations.  

Transit duration variations are thought to be one of the
few (but currently feasible) promising ways to find moons around extrasolar planets
\citep{exomoon1}, as the perturbing effect of a moon would alter both
the time of center transit and the duration of said transit for a
transiting planet. Secular TDVs can also be used to constrain the oblateness
of the central body, which has been done observationally for the
KOI-13 system \citep{findj2}. In this context, the stellar
oblateness leads to precession of the orbital elements and thereby
mimics the effects of secular interactions among multiple planets
(see equation \ref{bmatrixdiag}, which depends on the stellar
oblateness $J_{2}$). In order for TDVs to be a useful method to detect exomoons or measure stellar oblateness, the amplitude due to these effects must be large compared to the intrinsic variation which we determine here. \textcolor{black}{We also note that 
TDVs are now being compiled from the Kepler data \citep{mazeh}, 
with more data expected in the near future.}
The time series $\tau_{T}(t)$ yields two useful measures: first, it yields the transit duration variation rate, which can be parameterized as $\delta \tau_{T,t}$, the change in duration per unit time (in Table 1, we parameterize this as as a variation per year. For example: a TDV of 1 sec yr$^{-1}$ would mean that over one year, the expected duration would change by one second, regardless of when or how frequently the transits occur). Second, it yields the duration variation per orbit, $\delta \tau_{T,n}$, which can be directly compared to the magnitude of other effects that can also cause TDVs.
Both of these measures provide useful constraints on the properties of the system: the yearly TDVs provide approximate limits for the signal due to secular interactions between planets only. The duration variation per orbit allows for a fit to a series of durations over time, where:
\begin{equation}
\tau_{T}(t) = \tau_{T}(0) + \delta \tau_{T,n}\ n
\label{pal}
\end{equation}
where $n$ is the number of orbits observed. If this is done, then variation accumulates as $(\Delta \tau_{T}) = \delta \tau_{T,n} \ n$ when $(\Delta \tau_{T})$ is the total change in duration over an extended baseline of time. In this case, when the time series contains $N$ independent measurements, the precision in fitting $\delta \tau_{T,n}$, as given in Equation (\ref{pal}), is increased. The uncertainty scales like $\sigma \propto N^{-3/2}$, with one factor of $N^{-1}$ being due to the number of observed transits, a factor of $N^{-1/2}$ being due to the independent nature of these observations \citep[as used in][]{pal}. In this way, a large number of transit duration measurements can better constrain the TDV per orbit than would be possible looking at yearly drift alone using two widely separated transits \citep[see Figure 3 in][which is the first example of observed long-period TDVs of the type we would see for secular interactions considered in this work]{findj2}.

\begin{figure} 
\centering
\includegraphics[width=3.2in]{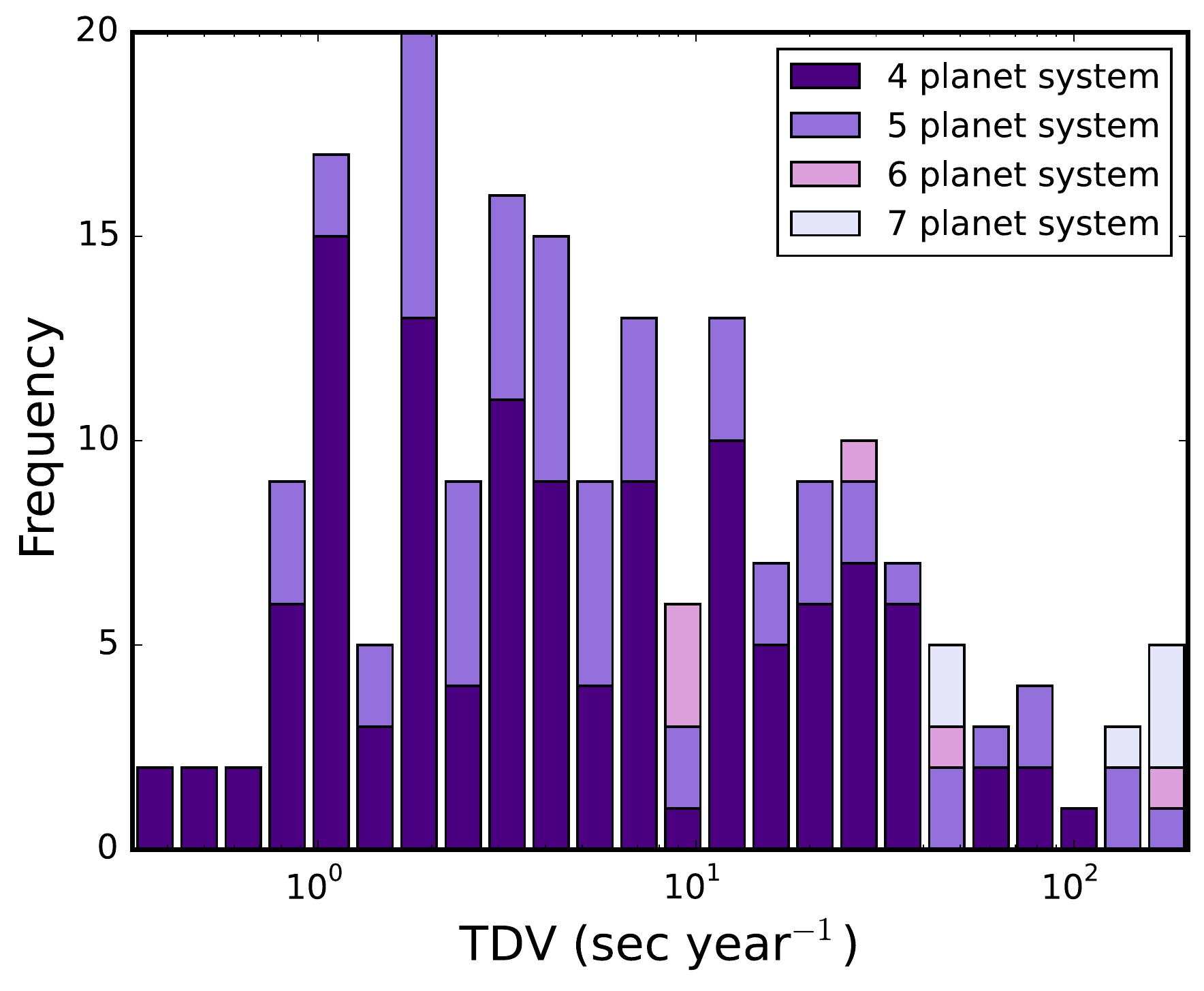} 
\includegraphics[width=3.2in]{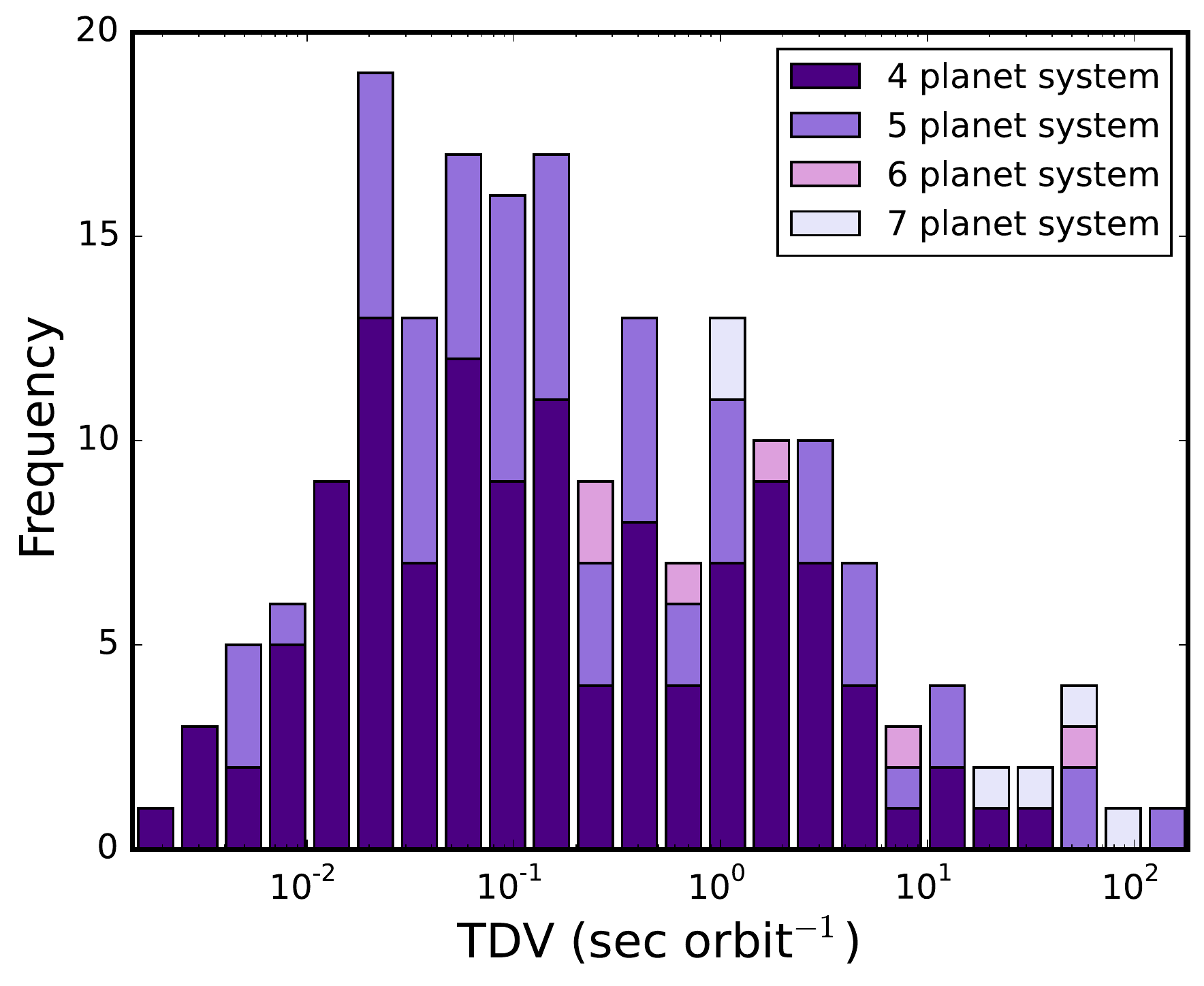} 

\caption{A histogram of the derived annual TDV values in this work, for Kepler systems with four or more coplanar planets. Data is presented in (\emph{upper panel}) TDV year$^{-1}$ and (\emph{lower panel}) TDV orbit$^{-1}$.
The bulk of the transit duration variations range from \textcolor{black}{0.01 to 10 seconds per orbit}. The data visualized here is also given in Table 2. This histogram includes only compact mutually transiting systems with four or more planets (Kepler-37, -48, and -68 are not included).  }
\label{fig:tdv_hist}
\end{figure}

The effect of secular interactions between planets in a multi-planet
system can occlude observations of other parameters traced by transit
durations (such as the presence of exomoons or solar oblateness), but
it can also provide evidence for additional planets in the system, as
non-transiting planets contribute to the duration variations even if
they are not directly observable.  

In Table 2, we present expected yearly TDVs for each planet considered in this
work. 
\textcolor{black}{These values are also presented in histograms in Figure \ref{fig:tdv_hist}.}
 Though these values are small because the yearly change in
inclination for each planet is very small, they provide limits for the
kind of TDVs expected in the observed Kepler multi-planet systems
without the presence of a perturber. The presence of a perturbing
secondary in any of these systems would lead to transit durations
outside the expected range. For example, circumbinary planets can exhibit TDVs on the order of hours \citep[such as for Kepler-47, as in][]{binaryplanets}. For exomoons, the TDV amplitude is expected to scale with $M_{s} a_{s}^{-1/2}$, when $s$ denotes a satellite \citep{exomoon1}. This amplitude is typically on the order of tens of seconds, being 13.7 seconds for the Earth-Moon system \citep{kippingbook}. In comparison, typical values for the secular interactions within a compact system are \textcolor{black}{a bit} smaller (being typically between $10^{-2} - 10^{1}$ seconds per orbit). 

\textcolor{black}{S}ignificant deviation in
transit durations above these \textcolor{black}{predicted} values would suggest the presence of an
additional effect (perturbing planet, extreme solar oblateness,
exomoon, etc.) in the system. The range of transit duration variations summarized in Figure \ref{fig:tdv_hist} thus serves as a baseline of the expected TDV distribution for tightly packed, coplanar, multi-planet systems.

\section{Planetary Mass Constraints}
\label{sec:mcs}
The observed current coplanarity of the Kepler multi-planet systems is a stringent constraint on the planets' orbital properties. 
For most of the planets in the Kepler system, the ratio $R_{p} /R_{*}$ is well-known. Combined with a value of the stellar radius (determined from either spectroscopy or interferometry), this value yields a measure of the planetary radius. 

To perform a dynamical analysis, these measured radii must be converted to mass. Although some Kepler planets have masses measured via long-term radial velocity surveys \citep{marcyrvsurvey}, the population of four-plus planet systems generally do not have measured masses due to the difficultly of measuring masses for small planets in multi-planet systems. Much recent work has been conducted aimed at finding a mass-radius relationship for exoplanets \citep{angie, lauren, leslie}. \textcolor{black}{When testing the CMT-stability of the compact Kepler systems, we use a supplemented version of the Wolfgang relation \citep{2015arXiv150407557W}. This relation introduces a large amount of scatter in density for planets that could be gaseous or rocky, which is useful for exploring the entire extent of parameter space in which the real planets could be living. However, another question that the apparent relative CMT-stability of the Kepler systems engenders is the effect of systematic mass enhancement \citep[which could be due to an incorrect measurement of the stellar radius, as in][in which the correction of such a misconception can be found]{koi961}. To test the effect of such systematic radius errors, we will inflate the masses of the constituent planets in the Kepler compact systems and examine the dynamical and CMT-stability of the systems. }

\textcolor{black}{
 For this experiment, we make a different choice in converting radii to masses: we use conversion law $M_P = M_{\earth} (R_P/R_{\earth})^{2.1}$ inferred from results of the Kepler mission \citep{lissauer}. Using this relation removes the scatter due to composition, enabling a qualitative study of the general stability status of the Kepler multi-planet systems, without noise from differing compositions between trials.}

\textcolor{black}{D}etermining the effect of planetary mass enhancement with respect to \textcolor{black}{roughly} estimated values would help determine if the parameter space of \textcolor{black}{CMT-stable} systems (which we have shown includes all the systems in our sample) changes if the planetary masses are systematically underestimated. 
To determine the extent of this parameter space, we evaluate the dynamical stability of the Kepler systems with varying mass enhancement factors, which places constraints on the maximum ratio by which the masses can be enhanced without losing the currently observing transiting configuration of the systems. 

To evaluate the effect of having larger planets in each system, we performed 40 numerical simulations of each system using \texttt{Mercury6} for each mass enhancement factor. The integration time for each system was $10^{6}$ dynamical times. This full treatment accounts for effects ignored in the secular theory such as the coupling of eccentricity and inclination, and instabilities due to orbit crossing or other effects. A mass enhancement factor describes the factor by which we increase all planetary masses within a single system. Although we alter the masses of the planets, we do not alter starting semi-major axes. The systems for each enhancement factor were created using observationally constrained orbital parameters supplemented with orbital parameters drawn from the standard priors (see Table \ref{table1}). When a system remains \textcolor{black}{CMT-stable} for the entire time, this means that it is observable in transit and the system as a whole does not go \textcolor{black}{dynamically} unstable (e.g., by ejecting a planet). 

There are two potential causes of instability in these systems. First, increased inclination oscillations can cause a some planets in a system to lie outside a mutual line of sight, even as a system remains dynamically stable. For the purposes of our analysis, we consider this to be an \textcolor{black}{CMT-}unstable system. Second, true dynamical instability (in the form of ejected/star-consumed planets or orbit crossing) also results in an \textcolor{black}{CMT-}unstable system. When either of these criteria (large inclination oscillations or true dynamical instability) is met for a certain mass enhancement factor, we categorize that system as unstable. 

We parameterize the dynamical fullness of a system in terms of the surface density of a disk consisting of the mass of its constituent planets spread over an annulus with an inner radius equal to the semi-major axis of the most interior planet, and an outer radius equal to the semi-major axis of the most exterior planet:
\begin{equation}
\Sigma = \frac{\sum_{i=1}^{i=n} m_{i}}{\pi (a_{n}^{2} - a_{1}^{2})}
\label{surfden}
\end{equation}
where $n$ is the number of planets in a system, $a$ is the semi-major axis, $m$ is the planetary mass, and $i$ denotes the planet number. 

\begin{figure} 
\centering
\includegraphics[width=3.2in]{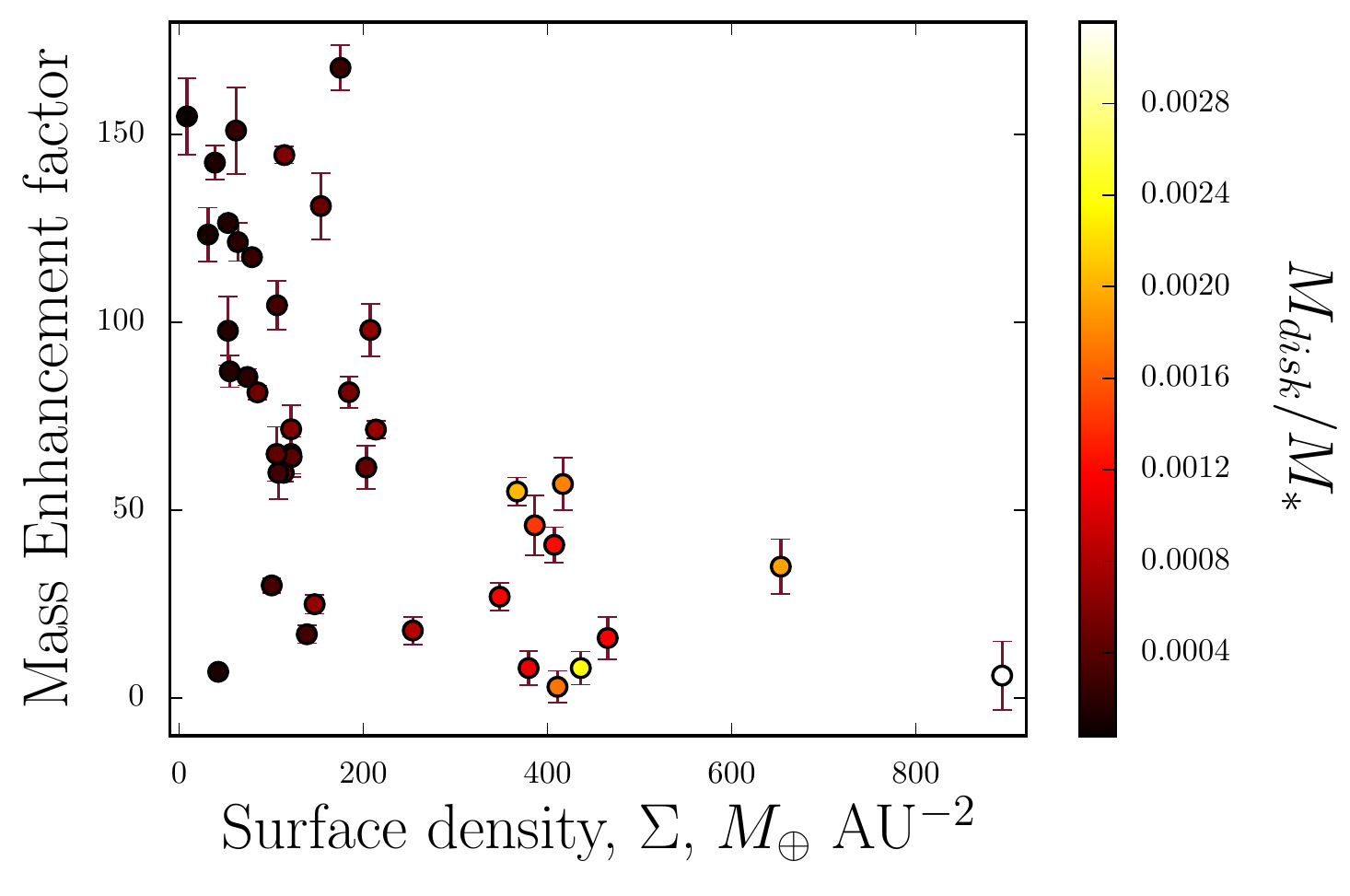} 
\caption{The mass enhancement factor required to knock a system out of a \textcolor{black}{CMT-}stable transiting configuration, plotted by the surface density of the annulus containing all the planets (which is defined in Equation \ref{surfden}). The points are shaded based on the ratio of the total planet mass to stellar mass ($M_{disk} / M_{*}$). The shape of the trend can be explained two ways. It could be explained by the existence of two distinct populations (one containing the disks where planet surface density is below 200 $M_{\earth} /$ AU$^{2}$, and another where the density is above 200 $M_{\earth} /$ AU$^{2}$, where the former are significantly less sensitive to mass enhancement), or it could be explained as a monotonic (but high-scatter)  decreasing trend with surface density. }
\label{fig:massdensity}
\end{figure}

In Figure \ref{fig:massdensity}, we plot the mass enhancement factor required to make a system \textcolor{black}{CMT-}unstable against the the surface density of the planet annuli. This plot is essentially a comparison of the dynamical fullness of the system (surface density) to the stability against excitation (mass enhancement factor required to knock a system out of transit). 
The observed result appears to intuitively support that a higher surface density of material leads to a less \textcolor{black}{CMT-}stable system (for which a lower mass enhancement factor can excited oscillations out of the plane). The large scatter of the data could also be explained by the existence of two distinct populations (one containing the disks where planet surface density is below 200 $M_{\earth} /$ AU$^{2}$, and another where the density is above 200 $M_{\earth} /$ AU$^{2}$, where the former are significantly less sensitive to mass enhancement).  

For many systems with a surface density $\Sigma >$ 200 $M_{\earth} /$ AU$^{2}$, hot or warm Jupiter-like planets would be \textcolor{black}{CMT-}stable even in a multiple-planet system. This finding suggests that Jovian-size planets can exist in tightly-packed multi-planet systems with semi-major axis similar to those of the discovered Kepler systems (although this result holds only for Myr timescales, as discussed below). 

The mass enhancement factor required to render the systems \textcolor{black}{CMT-}unstable may seem higher than expected. On one hand, the integrations are carried out for only $10^{6}$ dynamical times, which generally works out to be a few million years, which is short compared to the system ages. The critical enhancement factor appropriate for the ages of the systems are thus lower, but we assume here that the short-time values provide a good relative measure of stability. On the other hand, these systems are in \textcolor{black}{CMT-}stable configurations, even though their surface densities are much larger than that of out solar system (the analogous value for our solar system is 0.49 $M_{\earth} /$ AU$^{2}$). For comparison, we note that the GJ 876 system (one of the most dynamically active systems discovered to date) has a surface density $\Sigma = 2750\ M_{\earth} / $ AU$^{-2}$, which is much larger than the systems considered here.

\section{Conclusions} 
\label{sec:conclude} 

This paper has explored the dynamics of compact solar systems
undergoing oscillations in their orbital inclination angles. If such
oscillations occur with sufficient amplitude, then not all of the
planets in a multi-planet system are expected to transit at a given
epoch. By comparing the conditions required for the excitation of
inclination angles with the observed properties of compact
multi-planet systems, we can put constraints on their dynamical
history. In this work, we have provided measures of $\Delta b(t)$, the spread in impact parameters, and characterized the potential dynamical history \textcolor{black}{of} compact extrasolar systems. We have also utilized our method to test the dynamical \textcolor{black}{and CMT-}stability of a small sample of systems with additional non-transiting planets. From our derived $\Delta b(t)$, we have extracted subsequently the expected TDVs for observed systems in the case
that these systems have no additional non-transiting companions. Finally, we have explored the effect of enhancing the mass of planets in these tightly packed systems, with an aim at determining how robustly the transit stability holds as planetary masses increase. 

We have done this analysis by examining the multi-planet Kepler systems with
the greatest number of transiting planets and analyzing their
long-term stability, using a combination of secular (Sections \ref{sec:selfexcite}, \ref{sec:nontransit}, and \ref{sec:tranduration}) and numerical techniques (Sections \ref{sec:numerical} and \ref{sec:mcs}). Using the Kepler systems with the greatest number of transiting planets as our sample, we derived $\Delta b(t)$ for each planet using Monte Carlo techniques to marginalize over potential values of present orbital elements. We have determined that the compact Kepler
systems are \textcolor{black}{CMT-}stable against being excited into non-mutually-transiting
configurations.

Compact solar systems could have configurations that allow for a
significant spread in the orbital inclinations through secular
interactions between the constituent planets (Section
\ref{sec:selfexcite}). However, for the types of architectures
observed in the Kepler sample of multi-planet systems, the expected
range of inclination angles is almost always small. As shown in Figure 3, the
typical spread in the mean mutual impact parameter is typically less
than $\sim0.5$, whereas impact parameters greater than 2 are required
for planets to move out of transit. This result can also be expressed
in terms of inclination angles: self-excitation generally produces
$\Delta i \lta 0.5^\circ$, whereas angles of 1 -- 2$^\circ$ are
required to compromise transit in these compact systems. As a result, for most of the systems
discovered by the Kepler mission, the self-excitation of inclination
angle oscillations is generally not large enough to prevent planets
from being observed in transit.

\textcolor{black}{
We have also tested the behavior of generalized Kepler systems. For these generalizations, we drew orbital parameters for each system from expanded but observationally inspired posteriors, then tested the dynamical stability. For dynamically stable analogs, we proceeded with the analysis used for the observed Kepler systems. We found that the generalized systems are experience significantly more action in mutual impact parameter excitation, resulting in these systems being on average less CMT-stable than the observed Kepler systems. The observed Kepler systems are remarkably CMT-stable, even compared to their analogs.}

\textcolor{black}{Our derived result that self-excitation of inclination
angle oscillations is generally not large enough to prevent planets
from being observed in transit holds for the Kepler systems, but not their analogs; even then, it} has an important exception. We have also considered another type of Kepler system that contains 2
or 3 transiting planets and an additional planet not seen in transit
(where the additional body was discovered by radial velocity
follow-up).  Kepler 48 and Kepler 68 are examples of this type of system.
These systems are \textcolor{black}{CMT-}unstable to significant oscillations in inclination
angle, so that the expected spread in inclination angle is generally
large enough to move planets out of transit.  We found this result by
secularly evolving these systems after starting them in a nearly
coplanar configuration. Even starting roughly coplanar, the magnitude of these systems' self-excitation is large enough that not all planets can be seen in transit simultaneously for most of each system's orbital history. This finding indicates that the current Kepler systems with
non-transiting companions could have started roughly coplanar and subsequently had some of their planets
excited out of the plane via dynamical interactions between the planets that we know about. Specifically, it is not necessary to introduce additional bodies into these systems to recreate the currently observed architectures. 

We have focused on the secular interactions of compact systems of
planets, and derived observables corresponding to the current known
properties of these systems. These observables, the transit duration
variations for Kepler systems with the observationally determined
properties, are given in Table 2. Implicit in the motivation behind the
calculation of these TDVs is the idea that there could be additional
bodies in the systems we are considering, leading to true TDVs
deviating from those that we have found here. An additional massive
companion or an exomoon, for example, could cause transit duration variations with a larger amplitude than those derived in this work. If future
observations of TDVs in these systems are vastly different than
expected, it could potentially be evidence for either an exomoon or additional, exterior,
non-transiting bodies in these compact systems. 

We have also explored the effect of planetary mass enhancement in these systems. The stability of systems is related to how much the constituent planets' masses must be enhanced to result in a system that will no longer mutually transit. Generally, systems with higher effective surface density (calculated by spreading the mass of discovered planets within an annulus with inner and outer radii equal to the inner and outer planet's orbital radii) do not allow mass enhancement factors as high as those with lower surface density. This result suggests that dynamically `full' systems would not be mutually transiting if they hosted Jovian-mass planets. However, some systems with
lower surface densities would be \textcolor{black}{CMT-}stable in a transiting configuration
even with Jovian-mass planets (at least over time scales of $\sim10$
Myr), indicating that it might be possible to see multi-transiting
compact systems with Jovian-mass planets if they existed.
The stability boundaries -- over longer 
time scales -- should be explored further in future work.

Spreads in the inclination angles in compact systems can be
produced by a variety of astronomical processes, in addition to those considered in this work. Excitation by the
compact solar system planets themselves (with semi-major axes
$a\lta0.5$ AU) is not generally a significant effect, but we have not (yet)
calculated the effect caused by possible additional bodies in the
outer part of the solar system (where $a\approx5-30$ AU). Since planet
formation is a relatively efficient process, the additional giant
planets, not seen in transit by Kepler, are not only possible but
likely. The orbits of these outer planets can be endowed with high
inclination angles through a variety of dynamical mechanisms. For
example, most solar systems form within clusters, and inclinations can
be excited through dynamical interactions between solar systems and
other cluster members \citep{adams2001,malmberg2007,adams2010,gdawg}.  In
addition, a range of inclination angles can be realized through the
formation of planets in warped disks. The observed angular momentum
vectors in star-forming cores do not point in the same direction as a
function of radius \citep{goodman,caselli}.  This heterogeneity can
lead to differences in angular momentum vector of the disk plane as a
function of radius (for disks produced through collapse of the cores),
which in turn will influence the inclination angles of forming
planets \citep[see also][]{spalding}. These various mechanisms can lead to inclined, massive, outer
secondaries to the compact systems that we have considered in this
work.  The presence of such secondaries would alter the stability of
these systems, and this effect could be evident in the TDVs.
\textcolor{black}{Additionally, it is possible that a system of planets would have only some planets mutually transiting, instead of the condition of all planets in a system transiting that we have considered in this work. For example, although we see four planets in a system discovered by Kepler, it is possible that another short-period companions exists in such a system, resulting in our picture of the system being incomplete. Extensions on our calculation that account for this possibility could potentially be explored by using techniques such as the semi-analytical code CORBITS (Brakensiek \& Ragozzine, 2015).}

In summary, we have determined that self-excitation is not usually a dominant mechanism in
exciting mutual inclination in tightly packed, multi-planet
systems. Self-excitation does operate in some solar system architectures, where Kepler-48 and Kepler-68 are prime examples. Subsequent analysis of the effect of perturbing secondaries
and stellar fly-bys in a dense cluster environment will complete the
picture of how and when mutual inclinations are excited in
exoplanetary systems.

\section*{Acknowledgements}

We would like to thank Konstantin Batygin, Kathryn Volk, Ben Montet, Andrew
Vanderburg, and 
\textcolor{black}{Doug Lin} 
for useful conversations. We would like to additionally thank Konstantin Batygin for his careful review of the manuscript and helpful suggestions.  
\textcolor{black}{We would like to thank the referee, Darin Ragozzine, for his thoughtful and helpful suggestions.}
 J.B. is supported by the
National Science Foundation Graduate Research Fellowship, Grant
No. DGE 1256260.

$\,$

\label{lastpage} 

\clearpage 

\begin{table*}
\begin{minipage}{160mm} 
\label{tdv_table}
\centerline{\bf Transit Duration Variations for Kepler Compact Systems}
\centerline{$\,$} 
\begin{tabular}{ccccccc}
\hline 
\hline
\textbf{Planet}  & \textbf{Orbital Period, days} & \textbf{$\tau_{T,n}$, TDV (s yr$^{-1}$)} &  \textbf{$\tau_{T,t}$, TDV (s orbit$^{-1}$)}  \\
\hline

Kepler	11		&	10.3039	&	9.322660729	&	0.263177435	\\
Kepler	11		&	13.0241	&	8.478858911	&	0.302546593	\\
Kepler	11		&	22.6845	&	27.49182915	&	1.708598352	\\
Kepler	11		&	31.9996	&	9.333648315	&	0.818282226	\\
Kepler	11		&	46.6888	&	47.07565441	&	6.021659764	\\
Kepler	11		&	118.3807	&	195.5615046	&	63.42659673	\\
Kepler	20		&	3.6961219	&	1.914759018	&	0.019389542	\\
Kepler	20		&	6.098493	&	9.46333942	&	0.158115368	\\
Kepler	20		&	10.854092	&	4.0055073	&	0.119112725	\\
Kepler	20		&	19.57706	&	50.15850045	&	2.690290337	\\
Kepler	20		&	77.61184	&	73.36444964	&	15.59986281	\\
Kepler	24		&	4.244384	&	0.743325382	&	0.008643722	\\
Kepler	24		&	8.1453	&	1.056218584	&	0.023570458	\\
Kepler	24		&	12.3335	&	1.232153241	&	0.041634964	\\
Kepler	24		&	18.998355	&	3.605506001	&	0.187667625	\\
Kepler	26		&	3.543919	&	1.859049714	&	0.018050196	\\
Kepler	26		&	12.2829	&	3.710922038	&	0.124879135	\\
Kepler	26		&	17.2513	&	6.100397395	&	0.28832818	\\
Kepler	26		&	46.827915	&	37.7182322	&	4.839085401	\\
Kepler	32		&	0.74296	&	2.028539272	&	0.004129106	\\
Kepler	32		&	2.896	&	0.980507587	&	0.007779589	\\
Kepler	32		&	5.90124	&	1.7046195	&	0.027559914	\\
Kepler	32		&	8.7522	&	3.609925168	&	0.08656106	\\
Kepler	32		&	22.7802	&	5.932320689	&	0.370245073	\\
Kepler	33		&	5.66793	&	4.848962172	&	0.075297474	\\
Kepler	33		&	13.17562	&	5.861904438	&	0.211600617	\\
Kepler	33		&	21.77596	&	2.870395649	&	0.171248276	\\
Kepler	33		&	31.7844	&	10.62434609	&	0.925173879	\\
Kepler	33		&	41.02902	&	10.88222417	&	1.223252036	\\
Kepler	49		&	2.576549	&	1.222377872	&	0.008628812	\\
Kepler	49		&	7.2037945	&	2.229154696	&	0.043995541	\\
Kepler	49		&	10.9129343	&	3.964566165	&	0.118534384	\\
Kepler	49		&	18.596108	&	16.67313555	&	0.84946693	\\
Kepler	55		&	2.211099	&	0.74535018	&	0.004515186	\\
Kepler	55		&	4.617534	&	2.075484138	&	0.026256489	\\
Kepler	55		&	10.198545	&	7.323822446	&	0.204636528	\\
Kepler	55		&	27.9481449	&	7.955549215	&	0.609158472	\\
Kepler	55		&	42.1516418	&	23.59027592	&	2.724298248	\\
Kepler	62		&	5.714932	&	6.558019889	&	0.102681199	\\
Kepler	62		&	12.4417	&	10.787751	&	0.367720443	\\
Kepler	62		&	18.16406	&	7.847335247	&	0.390519091	\\
Kepler	62		&	122.3874	&	126.4774292	&	42.40888689	\\
Kepler	62		&	267.291	&	200.7871286	&	147.0372394	\\
Kepler	79		&	13.4845	&	12.8033798	&	0.473005959	\\
Kepler	79		&	27.4029	&	28.46032201	&	2.136699611	\\
Kepler	79		&	52.0902	&	20.63076067	&	2.944275204	\\
Kepler	79		&	81.0659	&	89.08814913	&	19.78633147	\\
Kepler	80		&	3.072186	&	2.130009061	&	0.017928175	\\
Kepler	80		&	4.645387	&	1.943170254	&	0.024730898	\\
Kepler	80		&	7.053	&	1.714628021	&	0.03313225	\\
Kepler	80		&	9.522	&	1.971262692	&	0.051425653	\\
Kepler	82		&	2.382961	&	0.965129741	&	0.006301004	\\
Kepler	82		&	5.902206	&	2.222408585	&	0.035937297	\\
Kepler	82		&	26.444	&	17.76864363	&	1.287326061	\\
Kepler	82		&	51.538	&	10.89753388	&	1.538731784	\\
Kepler	84		&	4.224537	&	4.439153917	&	0.051379096	\\
Kepler	84		&	8.726	&	2.002291695	&	0.047868486	\\
Kepler	84		&	12.883	&	3.658674592	&	0.129136177	\\
Kepler	84		&	27.434389	&	10.50305215	&	0.789437858	\\
Kepler	84		&	44.552169	&	29.82330904	&	3.640255081	\\
Kepler	85		&	8.306	&	1.592408413	&	0.036237108	\\
Kepler	85		&	12.513	&	2.059758945	&	0.070613051	\\
Kepler	85		&	17.91323	&	15.62065426	&	0.766620199	\\
Kepler	85		&	25.216751	&	26.0037509	&	1.796520854	\\

\hline
\end{tabular}
\end{minipage} 
\end{table*}    

\begin{table*}
\begin{minipage}{165mm} 
\centerline{\bf Transit Duration Variations for Kepler Compact Systems (continued)}
\centerline{$\,$} 
\begin{tabular}{cccccc}
\hline 
\textbf{Planet}  & \textbf{Orbital Period, days} & \textbf{$\tau_{T,n}$, TDV (s yr$^{-1}$)} &  \textbf{$\tau_{T,t}$, TDV (s orbit$^{-1}$)} \\
\hline

Kepler	90		&	7.008151	&	44.64259007	&	0.857156198	\\
Kepler	90		&	8.719375	&	51.72347308	&	1.235606461	\\
Kepler	90		&	59.73667	&	122.8193078	&	20.10086701	\\
Kepler	90		&	91.93913	&	163.2018666	&	41.10859624	\\
Kepler	90		&	124.9144	&	196.2255593	&	67.15451508	\\
Kepler	90		&	210.60697	&	160.6239691	&	92.6808971	\\
Kepler	90		&	331.60059	&	200.873026	&	182.492093	\\
Kepler	102		&	5.28696	&	21.67804261	&	0.314002587	\\
Kepler	102		&	7.07142	&	5.624323905	&	0.108964265	\\
Kepler	102		&	10.3117	&	3.349150204	&	0.094617622	\\
Kepler	102		&	16.1457	&	0.736673755	&	0.032586612	\\
Kepler	102		&	27.4536	&	42.68179251	&	3.210325641	\\
Kepler	106		&	6.16486	&	3.817466082	&	0.064477107	\\
Kepler	106		&	13.5708	&	7.56602341	&	0.281306823	\\
Kepler	106		&	23.9802	&	35.33153826	&	2.321253024	\\
Kepler	106		&	43.8445	&	18.90988445	&	2.271491586	\\
Kepler	107		&	3.179997	&	0.746863983	&	0.006506918	\\
Kepler	107		&	4.901425	&	0.979622681	&	0.013154924	\\
Kepler	107		&	7.958203	&	4.806241849	&	0.104791913	\\
Kepler	107		&	14.749049	&	1.19795885	&	0.048407545	\\
Kepler	122		&	5.766193	&	3.958098992	&	0.062529213	\\
Kepler	122		&	12.465988	&	0.796688261	&	0.027209606	\\
Kepler	122		&	21.587475	&	24.6377173	&	1.457167415	\\
Kepler	122		&	37.993273	&	64.29884982	&	6.692941794	\\
Kepler	150		&	3.428054	&	1.867968128	&	0.017543824	\\
Kepler	150		&	7.381998	&	1.135021003	&	0.022955405	\\
Kepler	150		&	12.56093	&	4.078868983	&	0.140368186	\\
Kepler	150		&	30.826557	&	18.32964457	&	1.548054337	\\
Kepler	169		&	3.250619	&	2.003755495	&	0.017845057	\\
Kepler	169		&	6.195469	&	3.473185364	&	0.058953458	\\
Kepler	169		&	8.348125	&	4.542706201	&	0.103898847	\\
Kepler	169		&	13.767102	&	4.044452569	&	0.152549017	\\
Kepler	169		&	87.090195	&	21.57958002	&	5.148958444	\\
Kepler	172		&	2.940309	&	0.46849388	&	0.003774019	\\
Kepler	172		&	6.388996	&	0.965003916	&	0.016891524	\\
Kepler	172		&	14.627119	&	3.850997739	&	0.154326033	\\
Kepler	172		&	35.118736	&	8.863030127	&	0.852762781	\\
Kepler	186		&	3.8867907	&	2.174369234	&	0.023154296	\\
Kepler	186		&	7.267302	&	2.671048208	&	0.053181682	\\
Kepler	186		&	13.342996	&	5.6398171	&	0.206170019	\\
Kepler	186		&	22.407704	&	21.23861165	&	1.303858968	\\
Kepler	186		&	129.9441	&	127.2638103	&	45.30734602	\\
Kepler	197		&	5.599308	&	1.746693303	&	0.026795271	\\
Kepler	197		&	10.349695	&	1.664512376	&	0.047197796	\\
Kepler	197		&	15.677563	&	2.946426313	&	0.126555573	\\
Kepler	197		&	25.209715	&	19.27829892	&	1.331508004	\\
Kepler	208		&	4.22864	&	0.327987695	&	0.003799841	\\
Kepler	208		&	7.466623	&	1.085357765	&	0.022202623	\\
Kepler	208		&	11.131786	&	2.145926971	&	0.065446575	\\
Kepler	208		&	16.259458	&	1.939023031	&	0.086376612	\\
Kepler	215		&	9.360672	&	5.214125094	&	0.133719767	\\
Kepler	215		&	14.667108	&	7.140540403	&	0.286934458	\\
Kepler	215		&	30.864423	&	13.54786835	&	1.145608602	\\
Kepler	215		&	68.16101	&	73.36673268	&	13.70068658	\\
Kepler	220		&	4.159807	&	1.08595549	&	0.012376343	\\
Kepler	220		&	9.034199	&	0.726348102	&	0.017978009	\\
Kepler	220		&	28.122397	&	31.65200923	&	2.438713341	\\
Kepler	220		&	45.902733	&	28.54622634	&	3.589999468	\\
Kepler	221		&	2.795906	&	1.622302263	&	0.012426862	\\
Kepler	221		&	5.690586	&	1.258209005	&	0.019616292	\\
Kepler	221		&	10.04156	&	3.116385606	&	0.085735269	\\
Kepler	221		&	18.369917	&	7.160429778	&	0.360373975	\\

\hline
\hline
\end{tabular}
\vspace{0.25cm}
\end{minipage} 
\end{table*}    

\begin{table*}
\begin{minipage}{165mm} 
\centerline{\bf Transit Duration Variations for Kepler Compact Systems (continued)}
\centerline{$\,$} 
\begin{tabular}{cccccc}
\hline 
\textbf{Planet}  & \textbf{Orbital Period, days} & \textbf{$\tau_{T,n}$, TDV (s yr$^{-1}$)} &  \textbf{$\tau_{T,t}$, TDV (s orbit$^{-1}$)} \\
\hline
Kepler	223		&	7.384108	&	6.794688993	&	0.1374595	\\
Kepler	223		&	9.848183	&	4.741440907	&	0.12793035	\\
Kepler	223		&	14.788759	&	3.494873231	&	0.141602296	\\
Kepler	223		&	19.721734	&	11.9376658	&	0.645017725	\\
Kepler	224		&	3.132924	&	1.856809099	&	0.015937649	\\
Kepler	224		&	5.925003	&	0.971515147	&	0.015770494	\\
Kepler	224		&	11.349393	&	3.646370338	&	0.113381068	\\
Kepler	224		&	18.643577	&	13.12338292	&	0.67032	\\
Kepler	235		&	3.340222	&	0.564712568	&	0.00516785	\\
Kepler	235		&	7.824904	&	7.345399102	&	0.15747135	\\
Kepler	235		&	20.060548	&	6.816955222	&	0.374662623	\\
Kepler	235		&	46.183669	&	39.50423791	&	4.998494925	\\
Kepler	238		&	2.090876	&	0.884749492	&	0.005068223	\\
Kepler	238		&	6.155557	&	2.172310058	&	0.036635009	\\
Kepler	238		&	13.233549	&	3.287882421	&	0.119206447	\\
Kepler	238		&	23.654	&	1.177061595	&	0.076280041	\\
Kepler	238		&	50.447	&	39.55458755	&	5.466877474	\\
Kepler	251		&	4.790936	&	3.555780556	&	0.04667265	\\
Kepler	251		&	16.514043	&	6.199763248	&	0.2805018	\\
Kepler	251		&	30.133001	&	10.65335888	&	0.879500476	\\
Kepler	251		&	99.640161	&	96.79454965	&	26.4236288	\\
Kepler	256		&	1.620493	&	0.323635851	&	0.001436848	\\
Kepler	256		&	3.38802	&	0.427493436	&	0.003968099	\\
Kepler	256		&	5.839172	&	0.884567462	&	0.014151073	\\
Kepler	256		&	10.681572	&	2.074977703	&	0.060723353	\\
Kepler	265		&	6.846262	&	3.529407786	&	0.066200686	\\
Kepler	265		&	17.028937	&	6.886638378	&	0.32129351	\\
Kepler	265		&	43.130617	&	31.42320459	&	3.713156719	\\
Kepler	265		&	67.831024	&	60.40531736	&	11.22562885	\\
Kepler	282		&	9.220524	&	15.78120361	&	0.398660183	\\
Kepler	282		&	13.638723	&	20.28678008	&	0.758043217	\\
Kepler	282		&	24.806	&	25.75148128	&	1.750112999	\\
Kepler	282		&	44.347	&	26.5277244	&	3.223082175	\\
Kepler	286		&	1.796302	&	0.713292306	&	0.003510379	\\
Kepler	286		&	3.468095	&	1.26801456	&	0.012048205	\\
Kepler	286		&	5.914323	&	3.526176454	&	0.05713684	\\
Kepler	286		&	29.221289	&	13.69225381	&	1.09617892	\\
Kepler	296		&	5.841648	&	2.228023946	&	0.035658443	\\
Kepler	296		&	10.21457	&	15.68127047	&	0.438842287	\\
Kepler	296		&	19.850242	&	16.85583517	&	0.916691527	\\
Kepler	296		&	34.14204	&	64.98518257	&	6.078703295	\\
Kepler	296		&	63.336	&	82.07825957	&	14.24248945	\\
Kepler	299		&	2.927128	&	1.224694325	&	0.009821471	\\
Kepler	299		&	6.885875	&	1.810056081	&	0.034147452	\\
Kepler	299		&	15.054786	&	12.26546764	&	0.505901344	\\
Kepler	299		&	38.285489	&	27.25944274	&	2.859290672	\\
Kepler	306		&	4.646186	&	3.606915153	&	0.045913421	\\
Kepler	306		&	7.240193	&	3.698312183	&	0.073360257	\\
Kepler	306		&	17.326644	&	10.59158411	&	0.502785225	\\
Kepler	306		&	44.840975	&	39.39149661	&	4.839323602	\\
Kepler	338		&	9.341	&	3.640821013	&	0.093175093	\\
Kepler	338		&	13.726976	&	3.068493345	&	0.115400369	\\
Kepler	338		&	24.310856	&	7.040320215	&	0.468921126	\\
Kepler	338		&	44.431014	&	13.81305425	&	1.681446594	\\
Kepler	341		&	5.195528	&	1.217989712	&	0.017337259	\\
Kepler	341		&	8.01041	&	1.039299363	&	0.022808806	\\
Kepler	341		&	27.666313	&	22.75719988	&	1.724952917	\\
Kepler	341		&	42.473269	&	12.26030004	&	1.426671292	\\
Kepler	402		&	4.028751	&	1.081566673	&	0.01193798	\\
Kepler	402		&	6.124821	&	0.917643194	&	0.015398357	\\
Kepler	402		&	8.921099	&	3.453650264	&	0.084411934	\\
Kepler	402		&	11.242861	&	2.886419391	&	0.088908526	\\

\hline
\hline
\end{tabular}
\vspace{0.25cm}
\end{minipage} 
\end{table*}    

\begin{table*}
\begin{minipage}{165mm} 
\centerline{\bf Transit Duration Variations for Kepler Compact Systems (continued)}
\centerline{$\,$} 
\begin{tabular}{cccccc}
\hline 
\textbf{Planet}  & \textbf{Orbital Period, days} & \textbf{$\tau_{T,n}$, TDV (s yr$^{-1}$)} &  \textbf{$\tau_{T,t}$, TDV (s orbit$^{-1}$)} \\
\hline

Kepler	444		&	3.6001053	&	2.256350153	&	0.022255063	\\
Kepler	444		&	4.5458841	&	1.40080575	&	0.017446303	\\
Kepler	444		&	6.189392	&	1.916669517	&	0.032501422	\\
Kepler	444		&	7.743493	&	2.784776198	&	0.059079164	\\
Kepler	444		&	9.740486	&	1.368811474	&	0.036528463	\\

\hline
\hline
\end{tabular}
\vspace{0.25cm}
\caption{Predicted values of the transit duration variations (TDVs) for 
the current sample of Kepler compact systems \textcolor{black}{containing only the planets that have been discovered so far}. Duration variations are presented both per orbit as well as per year. \textcolor{black}{Errors are typically on the order of 1\% of reported values, but are not reported for brevity.}  } 
\end{minipage} 
\end{table*}

\end{document}